\documentclass[journal]{IEEEtran}
\IEEEoverridecommandlockouts

\usepackage[T1]{fontenc}
\usepackage{cite}
\usepackage{amsmath,amssymb,amsfonts}
\usepackage{graphicx}
\usepackage{textcomp}
\usepackage{xcolor}
\usepackage{stmaryrd}
\usepackage[ruled,linesnumbered]{algorithm2e}

\usepackage{algpseudocode}
\usepackage{mathrsfs}
\usepackage{bm}
\usepackage[nocomma]{optidef}
\usepackage{booktabs}
\usepackage{diagbox}
\usepackage{multirow}
\usepackage{multicol}
\usepackage{enumitem}
\usepackage{tabularx}
\usepackage{amsthm}
\newtheorem{remark}{Remark}

\usepackage[scaled=.75]{beramono}
\newcommand{\bpara}[1]		{\medskip \noindent {\bf #1}}

\usepackage{xspace}

\def\BibTeX{{\mathrm B\kern-.05em{\sc i\kern-.025em b}\kern-.08em
    T\kern-.1667em\lower.7ex\hbox{E}\kern-.125emX}}
\begin{document}

\title{Joint Transmit and Receive Beamforming Design in Full-Duplex Integrated Sensing and Communications
}

\author{Ziang Liu, \IEEEmembership{}
        Sundar Aditya,~\IEEEmembership{Member,~IEEE,}
        Hongyu Li,~\IEEEmembership{Student Member,~IEEE,}
        and~Bruno Clerckx,~\IEEEmembership{Fellow,~IEEE}
\thanks{The authors are with the Communications \& Signal Processing (CSP) Group at the Dept. of Electrical and Electronic Engg., Imperial College London, SW7 2AZ, UK. (e-mails:\{ziang.liu20, s.aditya, c.li21, b.clerckx\}@imperial.ac.uk).}
\thanks{}}

\maketitle


\begin{abstract}
Integrated sensing and communication (ISAC) has been envisioned as a solution to realize the sensing capability required for emerging applications in wireless networks.
For a mono-static ISAC transceiver, as signal transmission durations are typically much longer than the radar echo round-trip times, the radar returns are drowned by the strong residual self interference (SI) from the transmitter, despite adopting sufficient SI cancellation techniques before digital domain - a phenomenon termed the \emph{echo-miss} problem. A promising approach to tackle this problem involves the ISAC transceiver to be full-duplex (FD), and in this paper we jointly design the transmit and receive beamformers at the transceiver, transmit precoder at the uplink user, and receive combiner at the downlink user to simultaneously (a) maximize the uplink and downlink communication rate, (b) maximize the transmit and receive radar beampattern power at the target, and (c) suppress the residual SI. To solve this optimization problem, we proposed a penalty-based iterative algorithm. Numerical results illustrate that the proposed design can effectively achieve up to 60 dB digital-domain SI cancellation, a higher average sum-rate, and more accurate radar parameter estimation compared with previous ISAC FD studies.
\end{abstract}


\section{Introduction}
Next generation wireless communication networks are expected to support high data transmission rates, high-quality wireless connectivity with massive devices, and highly accurate and robust sensing capability\cite{saad2019vision, liu2022integrated}. To realize these requirements, 
integrated sensing and communication (ISAC), which unifies the signal processing procedures and hardware framework between sensing and communication systems, is widely viewed as a promising solution to efficiently utilize the available spectral, hardware and energy resources.

\bpara{Challenge: \emph{Echo miss}.} In the existing ISAC literature, many studies assume that the sensing takes place in mono-static mode \cite{su2020secure, keskin2021limited, wu2022integrating, liu2022proximal, shi2022device} due to its relative simplicity compared to other sensing configurations (e.g., multi-static). In a mono-static ISAC system, the transmit (TX)/receive (RX) antennas are co-located, resulting in the transmit (dual-function) waveform being known at the receiver. Hence, the receiver can use the transmit waveform as a reference waveform to extract target information from the radar echo, thereby saving on the overhead associated with reference sharing.
However, the transmit waveform is also expected to serve communications users in parallel, and typically communication frames are much longer than the radar echo round-trip times (RTTs). For example, in the 5G NR specifications \cite{lin20195g}, a standard radio frame has $10$ms duration. For a target located at $100$-$1000$m from the radar, its echo RTT is of the order of $1$-$10\mu$s - orders of magnitude smaller than even the minimum unit of data scheduling (i.e., 1 slot  = 0.5ms). Hence, the radar echo is drowned by the severe self interference (SI) from the transmitter, which causes receiver saturation, where the power of the received signal exceeds the analog-to-digital converter (ADC) dynamic range. Even if there is no ADC saturation due to the use of sufficient SI cancellation techniques before quantization, the radar echo may still be difficult to detect because it is masked by strong residual SI. We term this phenomenon the \emph{echo-miss} problem. Thus, it is important to sufficiently suppress the residual SI to manageable levels. 

To address the \emph{echo-miss} problem and suppress the SI, one straightforward method is to physically separate the TX and RX antennas. The measurement-based study \cite{everett2014passive} shows that limited isolation capability can be achieved by a combination of directional isolation, absorber, and cross-polarization. Specifically, in the experiments, a $35$cm separation between the TX and RX antennas, along with an absorber, was shown to realize $45$dB passive suppression. However, the power level of the self interference (SI) can be large (i.e., up to $100$dB larger than the receiver thermal noise floor \cite{sabharwal2014band, zhang2016full}). Thus, physical separation of TX and RX antennas may not entirely solve the \emph{echo-miss} problem.
Consequently, to integrate communication and sensing functions, the transceiver should work in the full-duplex (FD) mode to simultaneously transmit a dual-functional signal, receive the echo signal, and suppress the SI, caused by the leakage of the transmit signal to the receiver. Some attempts have been made for ISAC in the FD context, with \cite{xiao2022waveform} concentrating on waveform design by utilizing the waiting time of conventional pulsed radars to transmit communication signal in a single-antenna setup. For the multi-antenna case, \cite{he2022joint} jointly optimizes relay beamformer, receive filter, and transmit power of the radar for a FD ISAC relay system, wherein the residual SI is assumed to be cancelled in advance by SI cancellation techniques.  However, for the most part, the SI cancellation problem is underestimated by many ISAC studies \cite{keskin2021limited, wu2022integrating, shi2022device, sturm2011waveform, kumari2019adaptive, kumari2017ieee}, which assume either ideal isolation between TX and RX or rely on the radar-function-focused SI cancellation method in \cite{barneto2019full}. 


\bpara{Previous Approaches for SI Cancellation.}
\begin{figure*}[t]
	\begin{center}
		\includegraphics[width =0.85\textwidth]{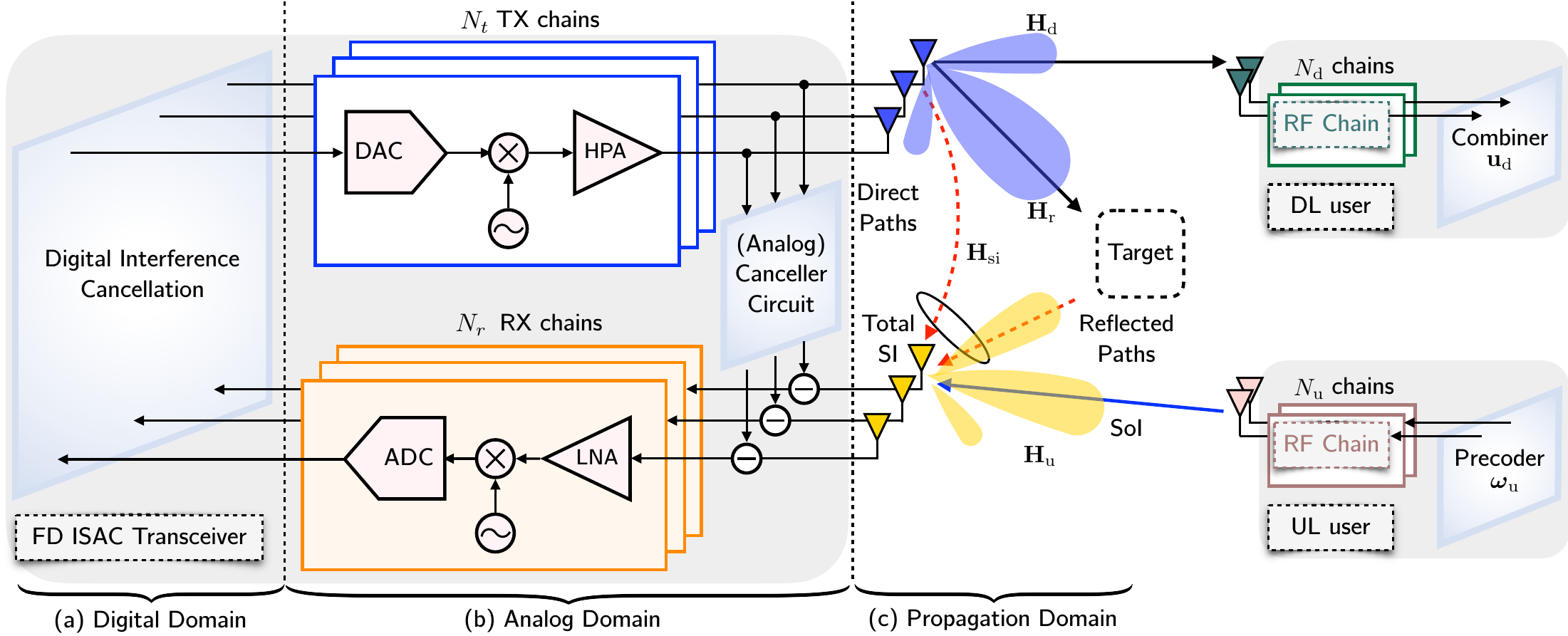}
		\caption{Full-duplex integrated sensing and communication system, and illustration of the boundaries and contents of the SI suppression in the propagation, analog, and digital domains for multi-antenna configurations.}
		\label{fig:sic3domain}
	\end{center}
\end{figure*}
The SI cancellation problem has been actively studied in FD wireless communication systems\cite{kolodziej2019band, duarte2012experiment, kolodziej2016multitap} (i.e., without the additional sensing functionality). In general, SI cancellation techniques can be adopted in the propagation \cite{everett2016softnull}, analog \cite{debaillie2014analog, korpi2016full, hong2014applications}, and digital domains \cite{ahmed2015all, komatsu2018basis}. As shown in Fig. \ref{fig:sic3domain}, the propagation-domain cancellation aims to minimize the coupling between the transmit and receive direct paths. This kind of cancellation is achieved by techniques based on path loss, cross-polarization, and antenna directionality \cite{everett2014passive}. Beyond this, the analog-domain cancellation aims to suppress SI before the ADC, where a negative copy of the transmit waveform generated by the canceller circuit is subtracted from the received signal \cite{debaillie2014analog, korpi2016full, kolodziej2016multitap, hwang2016digitally}. Finally, as the last defense against SI, the digital-domain cancellation block utilizes either linear or non-linear adaptive filters to generate the negative of the residual SI \cite{ahmed2015all, komatsu2018basis}, and add it to the digital signal after the ADC. 

The above SI cancellation techniques for communications rely on the uncorrelated nature between the SI (e.g., downlink data stream) and the signal of interest (SoI) (e.g., the uplink data stream); thus, the SI can be suppressed by adding its negative to the received signal without impairing the SoI. However, since the SoI in ISAC consists of uplink communications data and radar echoes that are correlated with the SI, it is challenging to effectively suppress the SI without distorting the radar echoes.
To tackle this problem, utilizing the time-of-arrival (ToA) difference or the spatial angle-of-arrival (AoA) difference between the SI and echo are two promising approaches. With respect to the first approach, the early study \cite{barneto2019full} utilizes the temporal difference to generate a negative counterpart of the SI signal before the ADC (cf. Fig. \ref{fig:sic3domain} (b)), based on a gradient-learning method. Apart from adding this negative counterpart, an adaptive filter is also employed to generate a negative in digital domain to cancel the residual SI (cf. Fig. \ref{fig:sic3domain} (a)). In practice, many factors (e.g., RF taps, adaptive filter taps, and update algorithms) affect the accuracy of the generated negatives in both domains, which in turn, have a huge impact on the SI cancellation performance. If the SI signal is fast-changing, this approach may fail in tracking and can have high computational complexity. 

In multi-antenna systems, an alternative way to suppress SI in ISAC is by employing the spatial AoA difference between the SI and SoI. In \cite{barneto2020beamforming}, the SI cancellation based on TX and RX beamforming design is adopted in the digital domain (cf. Fig. \ref{fig:sic3domain}a). Specifically, the TX beamformer is the weighted sum of two separate beams probing at a downlink communication user and a radar target, whose power allocation is controlled by a parameter. In terms of the RX beamformer, the null-space projection (NSP) based on pseudoinverse operation is used to generate nulls in the angles of the downlink communication beam and SI. In \cite{liyanaarachchi2021joint} and \cite{baquero2021beamformer}, the NSP method is utilized to design hybrid TX and RX beamformer for sensing the target and communicating to a downlink user (cf. Figs. \ref{fig:sic3domain}a and \ref{fig:sic3domain}b). However, in these studies, the TX and RX beamformers are separately designed and only the RX beamformer is used for SI cancellation. Recently, intelligent reflecting surfaces (IRS) have emerged as a means to boost the SoI, and thus reduce the effect of SI \cite{sharma2020intelligent, khel2022performance}, but this approach cannot actively cancel SI, and induces additional hardware and beamforming design complexity.

Thus, the potential of \emph{joint ISAC TX-RX beamformer design} at transceiver to further suppress the SI has not been explored. In addition, the uplink communication performance in the FD ISAC system has not been analyzed. Given that the research on FD ISAC is still in its infancy (cf. Table \ref{tab:1}), 
we consider a mono-static FD ISAC multiple input multiple output (MIMO) system. In this system, (a) the TX-RX beamformers at the transceiver, (b) transmit precoder at the uplink user, and (c) receive combiner at the downlink user are jointly optimized. The aim of the optimization is to simultaneously (a) maximize the uplink and downlink rate, (b) maximize the transmit and receive radar beampattern power at the target, and (c) suppress the residual SI. 
Inspired by research adopting the penalty-dual decomposition (PDD) method (e.g., for FD mmWave hybrid beamforming \cite{da20201}, and IRS \cite{zhao2021exploiting}), a penalty-based iterative algorithm is proposed to solve the optimization problem. Our proposed scheme can work when the received signal exceeds ADC dynamic range by adopting effective SI cancellation techniques before quantization.

\begin{table}[t]
  \centering

\caption{Novelty Comparison with Existing FD ISAC Literature}
\resizebox{0.5\textwidth}{!}{%
  \begin{tabular}{l|c c c c c}
    \toprule[1pt]
    & Our work   & \cite{xiao2022waveform} &  \cite{he2022joint} & \cite{barneto2019full} & \cite{barneto2020beamforming, liyanaarachchi2021joint, baquero2021beamformer}   \\ \hline \hline
    Relay                           &            &         & \checkmark &            &            \\ \hline
    Analog hardware design          &            &         &            & \checkmark &                  \\ \hline 
    Waveform design                 &            &\checkmark&            &            &                     \\ \hline
    Tx/Rx beamformers NSP design    &            &         &            &            & \checkmark        \\ \hline
    Tx/Rx beamformers joint design  & \checkmark &         & \checkmark &            \\ \hline
    Uplink user                     & \checkmark &         & \checkmark &            &            \\ \hline
    Downlink user                   & \checkmark &\checkmark& \checkmark &            & \checkmark \\ \hline
    Multi-antenna at users          & \checkmark &         &            &            &                     \\ \hline
    Radar performance               & \checkmark &\checkmark& \checkmark & \checkmark & \checkmark \\ \hline
    Communication performance       & \checkmark &\checkmark&            &            &                      \\
    
    \bottomrule[1pt]
  \end{tabular}
  \label{tab:1}
  }
\end{table}


    

\bpara{Contributions and Overview of Results.} In this paper, our contributions are summarized as follows:
\begin{itemize}
    \item We first model a FD ISAC mono-static system to capture the \emph{echo-miss} problem.
    
    \item To suppress the residual SI and preserve the two types of SoI (e.g., radar echo and uplink data), we formulate the joint TX-RX beamformer design problem for FD ISAC, where the objective function incorporates (a) uplink and downlink rates as the communications metric, (b) the transmit and receive radar beampattern power at a target as the sensing metric, and (c) the post-beamforming SI residual as a penalty term. Based on the equivalence between the rate maximization and the mean square error (MSE) minimization \cite{christensen2008weighted, shi2011iteratively}, and inspired by the PDD method, an iterative algorithm with guaranteed convergence and acceptable complexity is developed using block coordination descent (BCD) methods. As seen in Table \ref{tab:1}, in contrast to the existing literature, our optimization framework concentrates on joint TX-RX beamformer design and directly cancel residual SI.
    
    \item The performance of the designed beamformers is validated via simulations, which show that up to $60$dB residual SI can be effectively suppressed with better sum-rates for the uplink and downlink users and better radar parameter estimation performance (i.e., range-velocity and AoA detection) compared with the NSP method \cite{barneto2020beamforming, liyanaarachchi2021joint, baquero2021beamformer}. As shown in Table \ref{tab:1}, the performance of uplink communications, in particular, has not been thoroughly investigated in FD ISAC literature.

\end{itemize}

\bpara{Organization of This Paper.} The rest of the paper is organized as follows. The FD ISAC system model and the optimization framework for the joint ISAC TX-RX beamformers design is introduced in Section \ref{sec:system_model}. In Section \ref{sec:bf_design}, the problem reformulation and the proposed joint ISAC TX-RX beamformers design algorithm are provided based on the BCD method. The convergence and complexity analysis of the proposed algorithm is given in Section \ref{sec:analysis}, and numerical evaluations are presented in Section \ref{sec:result}. Finally, we conclude this work in Section \ref{sec:con}. 

\bpara{Notation.} The set of reals, integers, and complex numbers are denoted by $\mathbb{R}$, $\mathbb{Z}$, and $\mathbb{C}$, respectively. $\Re(x)$ and $\Im(x)$ denote the real and imaginary part of $x \in \mathbb{C}$, respectively. Continuous signals and discrete sequences are expressed by $x(t), t \in \mathbb{R}$ and $x[k], k \in \mathbb{Z}$, respectively. Matrices, vectors and scalars are written in capital boldface, small boldface and normal fonts, respectively. $[\mathbf{X}]_{i, :}$ and $[\mathbf{X}]_{:, j}$ denote the $i$-th row and $j$-th column of the matrix $\mathbf{X}$. $[\mathbf{X}]_{i,j}$ denotes the entry of the matrix $\mathbf{X}$ at index $(i, j)$. Similarly, $[\mathbf{x}]_{i}$ for vector $\mathbf{x}$. $\mathbf{X}^H$, $\mathbf{X}^\top$ and $\mathbf{X}^\dagger$ are used to denote conjugate-transpose, transpose and pseudo inverse of matrix $\mathbf{X}$, respectively.  We use $\mathscr{E}{(\cdot)}$, $|\cdot|$, and $\|\cdot\|_{2}$ to denote statistical expectation, absolute value and Euclidean norm.

\section{System Model}
\label{sec:system_model}
\subsection{Signal Model}
As shown in Fig. \ref{fig:sic3domain}, we consider a single-cell narrowband FD MIMO ISAC transceiver equipped with $N_\mathrm{t}$ transmit antennas and $N_\mathrm{r}$ receive antennas. All antenna arrays are assumed to be uniform linear arrays (ULA) with half-wavelength spacing between adjacent antenna elements. The transceiver simultaneously serves one uplink user with $N_\mathrm{u}$ antennas, one downlink user with $N_\mathrm{d}$ antennas, and probes a target direction. 


Let $s_\mathrm{d} \in \mathbb{C}$ denote the ISAC downlink transmit symbol, and $s_\mathrm{u} \in \mathbb{C}$ the uplink symbol. We assume both $s_\mathrm{d}$ and $s_\mathrm{u}$ have unit power. The received signal $\mathbf{y}_\mathrm{d} \in \mathbb{C}^{N_\mathrm{d} \times 1}$ at the downlink user is given by    
\begin{equation}
    \mathbf{y}_\mathrm{d} = {\mathbf{H}_\mathrm{d}} \mathbf{p} s_\mathrm{d} + \mathbf{n_\mathrm{d}},
    \label{eq:ydl}
\end{equation}
where $\mathbf{p} \in \mathbb{C}^{N_\mathrm{t} \times 1}$ denotes the transmit precoder at the transceiver,
${\mathbf{H}_\mathrm{d}} \in \mathbb{C}^{N_\mathrm{d} \times N_\mathrm{t}}$ the downlink communication channel, $\mathbf{n_\mathrm{d}} \in \mathbb{C}^{N_\mathrm{d} \times 1}$ the independent and identically distributed (i.i.d) additive complex Gaussian noise (i.e.,  $\mathbf{n}_\mathrm{d} \sim \mathcal{C} \mathcal{N} (\mathbf{0}, \sigma^{2}_{\mathrm{d}} \mathbf{I}_{N_{\mathrm{d}}} )$). 
At the downlink user, an estimate of ${s}_\mathrm{d}$, denoted by $\widehat{s}_\mathrm{d}$, is obtained by filtering $\mathbf{y}_\mathrm{d}$ by a combiner $ \mathbf{u}_\mathrm{d} \in \mathbb{C}^{N_\mathrm{d} \times 1}$, as follows:
\begin{equation}
    \widehat{s}_\mathrm{d} = \mathbf{u}_\mathrm{d}^{H} \mathbf{y}_\mathrm{d} = \mathbf{u}_\mathrm{d}^{H} {\mathbf{H}_\mathrm{d}} \mathbf{p} s_\mathrm{d} + \mathbf{u}_\mathrm{d}^H \mathbf{n}_\mathrm{d}.
    \label{eq:signaldl}
\end{equation}

At the FD ISAC transceiver, the received signal incorporates the SoI (i.e., the radar echo and the uplink symbol, $s_\mathrm{u}$), along with residual SI. We assume that the received signal has no clipping error due to sufficient SI cancellation techniques in propagation and analog domains, and a high ADC dynamic range\footnote{The dynamic range is defined by the ratio between the largest and smallest possible values of the input signal, which are respectively the residual SI and radar echo in our system model. From our link budget simulation, the ratio between the residual SI and radar echo is $134.15$dB, thus we assume that the SI cancellation techniques before digital domain can achieve larger than $54.15$dB SI cancellation.}, of the order of $80$dB, which can be achieved by a variety of ways, e.g., logarithmic ADC\cite{lee20092}, modulo ADC \cite{liu2022lambda, bhandari2020unlimited}). Hence, the received signal at the transceiver is given by
\begin{equation}
        \mathbf{y}_\mathrm{u}  = \mathbf{H}_\mathrm{u} \boldsymbol{\omega}_\mathrm{u} s_\mathrm{u} + \mathbf{H} \mathbf{p} s_\mathrm{d}+\mathbf{n}_\mathrm{u},
    \label{eq:ybs}
\end{equation}
where $\boldsymbol{\omega}_\mathrm{u} \in \mathbb{C}^{N_\mathrm{u} \times 1}$ denotes the precoder vector of the uplink user, $\mathbf{H}_\mathrm{u} \in \mathbb{C}^{N_\mathrm{r} \times N_\mathrm{u}}$ the uplink communication channel, $\mathbf{H} \in \mathbb{C}^{N_\mathrm{r} \times N_\mathrm{t}}$ the aggregate channel comprising the radar channel, $\mathbf{H}_\mathrm{r} \in \mathbb{C}^{N_\mathrm{r} \times N_\mathrm{t}}$, and the self-interference channel, $\mathbf{H}_\mathrm{si} \in \mathbb{C}^{N_\mathrm{r} \times N_\mathrm{t}}$ (i.e., $\mathbf{H} = \mathbf{H}_\mathrm{r} + \mathbf{H}_\mathrm{si}$), and finally $\mathbf{n}_\mathrm{u} \sim \mathcal{C} \mathcal{N}\left(\mathbf{0},  \sigma^{2}_{\mathrm{u}} \mathbf{I}_{N_{\mathrm{r}}}\right) \in \mathbb{C}^{N_\mathrm{r} \times 1}$ the i.i.d additive complex Gaussian noise. The above-mentioned channels will be elaborated in Section \ref{sec:channel}.

At the FD ISAC transceiver, an estimate of ${s}_\mathrm{u}$, denoted by $\widehat{s}_\mathrm{u}$, is obtained by filtering $\mathbf{y}_\mathrm{u}$ by a receive beamformer $\mathbf{w} \in \mathbb{C}^{N_\mathrm{r} \times 1}$ as follows:
\begin{align}
    \widehat{s}_\mathrm{u} &= \mathbf{w}^H \mathbf{y_u} \notag \\
    &= \underbrace{\mathbf{w}^{H}  \mathbf{H}_\mathrm{u} \boldsymbol{\omega_\mathrm{u}} s_\mathrm{u}}_\text{Uplink stream}  
      + \underbrace{\mathbf{w}^{H} \mathbf{H}_\mathrm{r} \mathbf{p} s_\mathrm{d}}_\text{Radar return} + \underbrace{\mathbf{w}^H \mathbf{H}_\mathrm{si} \mathbf{p} s_\mathrm{d}}_\text{Residual SI} + \mathbf{w}^{H} \mathbf{n}_\mathrm{u},
    \label{eq:signalbs}
\end{align}
In Section \ref{sec:bf_design}, we design beamformers $\mathbf{w}$ and $\mathbf{p}$ such that $\mathbf{w}^H \mathbf{H}_\mathrm{si} \mathbf{p} \approx 0$ (i.e., SI suppression) while the uplink symbol and the radar returns are amplified. To decode $s_\mathrm{u}$, we treat the interfering radar return as noise, while we use the favourable correlation properties of communications signals \cite{aditya2022sensing} to isolate the uplink data stream from the radar returns to estimate the target parameters. We provide more details regarding this in Section \ref{sec:result}.

\begin{remark}
It is intuitive to choose a separate radar signal, and design two separate beamformers for radar and communications. However, for a fixed transmit power budget, this leads to the important question of power allocation across
the communication and sensing symbols. Hence, it is not obvious that radar performance
may be better with separate signals/precoders. Furthermore, for full-duplex operation,
beamformer design with separate radar and communication signals is more complex as
residual SI needs to be suppressed twice. For these reasons, we adopt the dual-functional
transmit signal and a common beamformer.
\end{remark}

\subsection{Channel Model}
\label{sec:channel}
\subsubsection{Communication Channel}
\label{sec:com_chan}
The communication channels, $\mathbf{H}_\mathrm{u}$ and $\mathbf{H}_\mathrm{d}$, are assumed to experience both small-scale and large-scale fading, which is modelled as follows, taking $\mathbf{H}_\mathrm{u}$ as an example
\begin{equation}
    \mathbf{H}_\mathrm{u} = \sqrt{\eta_\mathrm{u}} \, \mathbf{G}_\mathrm{u},
    \label{eq:com_chan_u}
\end{equation}
where $\eta_\mathrm{u} \in \mathbb{C}$ is the large-scale fading coefficient (including geometric attenuation and shadow fading), and $\mathbf{G}_\mathrm{u} \in \mathbb{C}^{N_\mathrm{r} \times N_\mathrm{u}}$ is the small-scale fading matrix modelled by the classic Rician fading model \cite{tse2005fundamentals}, as follows
\begin{equation}
    \mathbf{G}_\mathrm{u} = \sqrt{\frac{\kappa}{\kappa+1}} \bar{\mathbf{G}}_\mathrm{u} + \sqrt{\frac{1}{\kappa+1}} \tilde{\mathbf{G}}_\mathrm{u},
    \label{eq:rician_chan}
\end{equation}
where $\kappa \quad (\geq 0)$ captures the proportion of the energy in the direct path (LoS), relative to the energy of the scattered paths (NLoS). We assume that the size of the antenna arrays is negligible compared with the distance between transceiver and user. Thus, the channel matrix corresponding to the LoS path (denoted by $\bar{\mathbf{G}}_\mathrm{u}$ in \eqref{eq:rician_chan}) is given by
\begin{equation}
    \bar{\mathbf{G}}_\mathrm{u} = \mathbf{a}(\theta_\mathrm{u}) \mathbf{b}^\top(\theta^\prime_\mathrm{u}),
    \label{eq:los_row}
\end{equation}
where
\begin{enumerate}[leftmargin = *, label = ---]
    \item $\theta_\mathrm{u}$ and $\theta^{\prime}_\mathrm{u}$ denote the AoA at the transceiver and angle-of-departure (AoD) at the uplink user, respectively.
    \item $\mathbf{a}(\theta_\mathrm{u}) = \left[1, e^{j 2 \pi \delta \sin \left(\theta_{\mathrm{u}}\right)}, \ldots, e^{j 2 \pi\left(N_{\mathrm{r}}-1\right) \delta \sin \left(\theta_{\mathrm{u}}\right)}\right]^\top \in \mathbb{C}^{N_\mathrm{r} \times 1}$ is the steering vector in the direction of the AoA at the transceiver, with $\delta = 0.5$.
    \item $\mathbf{b}(\theta^\prime_\mathrm{u}) = \left[1, e^{j 2 \pi \delta \sin \left(\theta^\prime_{\mathrm{u}}\right)}, \ldots, e^{j 2 \pi\left(N_{\mathrm{u}}-1\right) \delta \sin \left(\theta^\prime_{\mathrm{u}}\right)}\right]^\top \in \mathbb{C}^{N_\mathrm{u} \times 1}$ is the steering vector in the direction of the AoD at the uplink user.
\end{enumerate}
Additionally, $\tilde{\mathbf{G}}_\mathrm{u}$ is the i.i.d. complex Gaussian matrix, where each element follows $[\tilde{\mathbf{G}}_\mathrm{u}]_{i.j} \sim \mathcal{CN}(0, 1)$.


\subsubsection{Radar Channel}
\label{subsubsec:radar}
We assume that the radar operates in track model, where an initial estimate of the target's AoA is available from previous scanning.
As in \eqref{eq:ybs}, the FD ISAC transceiver will receive a reflected echo with the target's AoA, range and velocity information (i.e., $\theta_\mathrm{r}$, $r$, $v$, respectively). Since we assume the transceiver to operate as a mono-static radar, the AoA and AoD are both equal to  $\theta_\mathrm{r}$. We model the radar channel as varying slowly with time, with its expression given by
\begin{equation}
    \mathbf{H}_\mathrm{r}(t) = \eta_\mathrm{r} e^{j 2 \pi f_\mathrm{d} t} \mathbf{a}(\theta_\mathrm{r}) \mathbf{b}^\top(\theta_\mathrm{r}),
    \label{eq:H_r}
\end{equation}
where $f_\mathrm{d} = 2 v f_\mathrm{c} /c$ is the Doppler frequency, $f_\mathrm{c}$ and $c$ are the signal carrier frequency and the speed of light. $\eta_\mathrm{r}$ is the attenuation factor, including round-trip path-loss and radar cross-section (RCS) $\sigma_{\mathrm{r}}$, given in \eqref{eq:radar_fad} derived from the radar range equation \cite{skolnik1980introduction}.
\begin{equation}
    \eta_\mathrm{r} = \sqrt{\frac{\lambda^{2} \sigma_{\mathrm{r}}}{(4 \pi)^{3} r^{4}}}.
    \label{eq:radar_fad}
\end{equation}

\subsubsection{Self-Interference Channel}
In this work, we focus on the residual digital-domain SI cancellation assuming that the analog SI cancellation method is effective, i.e., there is no saturation at the ADC of the receive RF chains at the transceiver.
To obtain $\mathbf{H}_\mathrm{si}$, channel estimation methods based on classic estimation algorithms (e.g., least square (LS) \cite{ahmed2015all}, maximum-likelihood (ML) \cite{masmoudi2015maximum}) or machine learning-based methods \cite{muranov2021deep} can be used. For illustration and without loss of generality, we consider a channel model derived from experiments \cite{duarte2012experiment, nguyen2014spectral}, which is similar to \eqref{eq:rician_chan} and given by:
\begin{equation}
    \mathbf{H}_\mathrm{si} = \sqrt{\eta_\mathrm{si}} \, \mathbf{G}_\mathrm{si} ,
    \label{eq:si_chan}
\end{equation}
where $\eta_\mathrm{si}$ and $\mathbf{G}_\mathrm{si}$ denote the large-scale fading coefficient and small-scale fading matrix, respectively. Specifically, $\mathbf{G}_\mathrm{si}$ has the same structure as $\mathbf{G}_\mathrm{u}$ in \eqref{eq:rician_chan}, containing LoS and NLoS components.


\subsection{Problem Formulation}
\subsubsection{Communication Rate}
With the uplink channel $\mathbf{H}_\mathrm{u}$, radar channel $\mathbf{H}_\mathrm{r}$\footnote{For notation simplicity, we drop the time instant, $(t)$, in $\mathbf{H}_{\mathrm{r}}(t)$.}, and SI channel $\mathbf{H}_\mathrm{si}$ in \eqref{eq:com_chan_u}, \eqref{eq:H_r}, and \eqref{eq:si_chan}, the uplink communication rate is given by
\begin{equation}
\label{eq:r_u}
    R_\mathrm{u} = \log_2(1+\gamma_\mathrm{u}),
\end{equation}
where $\gamma_\mathrm{u}$ is the Signal-to-Interference-plus-Noise Ratio (SINR) of the received uplink signal. The expression for $\gamma_\mathrm{u}$ is given by
\begin{equation}
    \gamma_\mathrm{u} = \frac{|\mathbf{w}^H \mathbf{H}_\mathrm{u} \boldsymbol{\omega}_\mathrm{u}|^2}{|\mathbf{w}^H \mathbf{H} \mathbf{p}|^2 + \| \mathbf{w} \|^2 \sigma_\mathrm{u}^2}.
    \label{eq:upsinr}
\end{equation}
Similarly, the downlink communication rate is given by
\begin{equation}
    R_\mathrm{d} = \log_2(1+\gamma_\mathrm{d}),
    \label{eq:r_d}
\end{equation}
where $\gamma_\mathrm{d}$ is the SINR of the downlink signal given by
\begin{equation}
    \gamma_\mathrm{d} = \frac{|\mathbf{u}_\mathrm{d}^H \mathbf{H}_\mathrm{d} \mathbf{p}|^2}{\|\mathbf{u}_\mathrm{d}\|^2 \sigma_\mathrm{d}^2}.
\end{equation}
In FD systems, in general, the downlink user would experience interference from uplink transmissions, as well. However, we assume that this interference is effectively managed by scheduling \cite{bai2013distributed, sahai2013degrees, karakus2015opportunistic}, and thus ignore its effect in \eqref{eq:r_d}.

\subsubsection{Radar Beampattern Power}
The transmit and receive beampattern power in the target direction, denoted by $G_\mathrm{t}$ and $G_\mathrm{r}$, respectively, as a function of $\mathbf{p}$ and $\mathbf{w}$, is given by
\begin{equation}
    G_\mathrm{t} = |\mathbf{b}^H(\theta_\mathrm{r}) \mathbf{p}|^2,  \quad  G_\mathrm{r} = |\mathbf{w}^H \mathbf{a}(\theta_\mathrm{r})|^2.
\end{equation}
\begin{remark}
We use beampattern power as a proxy for radar SINR. While the latter may be a more intuitive sensing metric, its fractional form makes it non-convex -- and thus, intractable -- as well as difficult to obtain good bounds. On the other hand, the tractability of the beampattern power metric enables us to obtain closed-form solutions for key sub-problems (see Section \ref{subsec:block}) that help Algorithm \ref{alg:alg1} converge faster.
\end{remark}


\subsubsection{Objective Function and Constraints}
We aim to design the transmit and receive beamformer $\mathbf{p}, \mathbf{w}$ at the transceiver, precoder $\boldsymbol{\omega}_\mathrm{u}$ at the uplink user, and combiner $\mathbf{u}_\mathrm{d}$ at the downlink user to simultaneously maximize the uplink and downlink rate, the transmit and receive radar beampattern power at the target, and suppress the residual SI. In addition, the following constraints should be satisfied: 1) transmit power constraint at the transceiver and uplink user, i.e. $\|\mathbf{p}\|^2 \leq P_\mathrm{d}$, $\|\boldsymbol{\omega}_\mathrm{u}\|^2 \leq P_\mathrm{u}$, where $P_\mathrm{d}$ and $P_\mathrm{u}$ are the maximum transmit power permitted at transceiver and uplink user, respectively; 2) receive power constraint at the transceiver, i.e. $\|\mathbf{w}\|^2 = 1$; 3) SI suppression constraint, i.e., $\mathbf{w}^H \mathbf{H}_\mathrm{si} \mathbf{p} = 0$. With the above constraints, we can formulate the joint ISAC TX-RX beamformers design as 
\begin{maxi!}|s|[2]                   
    {\mathbf{p}, \mathbf{w},\boldsymbol{\omega}_\mathrm{u},\mathbf{u}_\mathrm{d}}                               
    {\alpha_1 R_\mathrm{u} + \alpha_2 R_\mathrm{d} + \alpha_3 G_\mathrm{t} + \alpha_4 G_\mathrm{r} \label{eq:op1}}   
    {\label{eq:p1}}             
    {} 
    \addConstraint{\|\mathbf{p}\|^2_2}{\leq P_\mathrm{d}, \label{eq:op1c1}}
    \addConstraint{\|\boldsymbol{\omega}_\mathrm{u}\|^2_2}{\leq P_\mathrm{u}, \label{eq:op1c2}}
    \addConstraint{\|\mathbf{w}\|^2_2}{ =1, \label{eq:op1c3}}
    \addConstraint{\mathbf{w}^H \mathbf{H}_\mathrm{si} \mathbf{p}}{=0. \label{eq:op1c4}}
\end{maxi!}

The weight of the four components in \eqref{eq:op1} is controlled by the coefficient $\alpha_n, \forall n \in \{1, 2, 3, 4\}$. The problem \eqref{eq:p1} is difficult to solve due to the following reasons: 1) the coupling between the transmit and receive beamformer in constraint \eqref{eq:op1c4}; 2) the non-convex communication metrics (i.e., $R_\mathrm{u/d})$ in objective \eqref{eq:op1};
3) the quadratic radar metrics (i.e. $G_\mathrm{t/r})$ in objective \eqref{eq:op1} in the maximization problem. To solve \eqref{eq:p1}, we transform it into an alternative form that permits an iterative solution with guaranteed convergence to a feasible solution. 


\section{Joint TX-RX Beamformer Design}
\label{sec:bf_design}
\subsection{Problem Transformation}
\label{subsec:trans}

To address the difficulties in the previous section, we transform problem \eqref{eq:p1} based on the PDD method \cite{shi2020penalty}, the weighted Minimum Mean Square Error (MMSE) approach \cite{shi2011iteratively}, and the reformulation of the quadratic beampattern power function\cite{xu2020multi}. Each of these techniques addresses a specific difficulty, as explained below:

\subsubsection{Coupling Constraint Transformation by Penalty Term}
To tackle the coupling between $\mathbf{w}$ and $\mathbf{p}$ in constraint \eqref{eq:op1c4}, 
we transform the constraint $\mathbf{w}^H \mathbf{H}_\mathrm{si} \mathbf{p} = 0$ into a penalty term $\frac{1}{2\beta} \|\mathbf{w}^H \mathbf{H}_\mathrm{si} \mathbf{p} \|^2_2$, and add it to the objective function with the penalty parameter $\beta > 0$. In particular, $\beta$ determines the penalty intensity of the digital-domain SI suppression, with lower values of $\beta$ prioritizing SI suppression over communication and sensing performance. Thus, problem \eqref{eq:p1} can be reformulated as \eqref{eq:p2}.
\begin{subequations}
\label{eq:p2}
    \begin{align}
    &\underset{\substack{\{\mathbf{p}, \mathbf{w}\}, \{\boldsymbol{\omega}_\mathrm{u},\mathbf{u}_\mathrm{d}\}}}{\text{max}}
    && \alpha_1 R_\mathrm{u} + \alpha_2 R_\mathrm{d} + \alpha_3 G_\mathrm{t} + \alpha_4 G_\mathrm{r} \nonumber\\
    &&& -\frac{1}{2\beta} \|\mathbf{w}^H \mathbf{H}_\mathrm{si} \mathbf{p} \|^2_2 \label{eq:op2} \\
    &\qquad \text{s.t.}
    &&\eqref{eq:op1c1}-\eqref{eq:op1c3}. \nonumber
    \end{align}
\end{subequations}


\subsubsection{Communication Rate Transformation by WMMSE Method}
\label{sssec:wmmse}
To circumvent the non-convexity of the rate terms in \eqref{eq:op2}, we adopt the weighted minimize mean square error (WMMSE) method to reformulate these terms based on the equivalence between the rate and MSE matrix \cite{christensen2008weighted, shi2011iteratively}. With the received signal \eqref{eq:signalbs} at the transceiver, the MSE matrix at the transceiver is defined as
\begin{equation}
\begin{aligned}
    \mathsf{E}_\mathrm{BS} 
    & \triangleq \mathscr{E}\left\{\left(\widehat{\mathbf{s}}_\mathrm{u}-\mathbf{s}_\mathrm{u}\right)\left(\widehat{\mathbf{s}}_\mathrm{u}-\mathbf{s}_\mathrm{u}\right)^{H}\right\}\\
    & = \mathbf{w}^H \mathbf{H}_\mathrm{u} \boldsymbol{\omega}_\mathrm{u} \boldsymbol{\omega}_\mathrm{u}^H \mathbf{H}_\mathrm{u}^H \mathbf{w} - 2 \Re\{\mathbf{w}^H \mathbf{H}_\mathrm{u} \boldsymbol{\omega}_\mathrm{u} \} \\
    & \quad + \mathbf{w}^H (\mathbf{H} \mathbf{p} \mathbf{p}^H \mathbf{H}^H + \sigma^2_\mathrm{u}\mathbf{I}) \mathbf{w} +1.
    \label{eq:msebs}
\end{aligned}
\end{equation}
Similarly, MSE matrix function of downlink user can be also defined as 
\begin{equation}
\begin{aligned}
    \mathsf{E}_\mathrm{d} 
    & \triangleq \mathscr{E}\left\{\left(\widehat{\mathbf{s}}_\mathrm{d}-\mathbf{s}_\mathrm{d}\right)\left(\widehat{\mathbf{s}}_\mathrm{d}-\mathbf{s}_\mathrm{d}\right)^{H}\right\}\\
    & = \mathbf{u}^H_\mathrm{d} \mathbf{H}_\mathrm{d} \mathbf{p} \mathbf{p}^H \mathbf{H}^H_\mathrm{d} \mathbf{u}_\mathrm{d} - 2 \Re \{\mathbf{u}^H_\mathrm{d}  \mathbf{H}_\mathrm{d} \mathbf{p}\}\\
    & \quad + \sigma^2_\mathrm{d} \mathbf{u}^H_\mathrm{d} \mathbf{u}_\mathrm{d} + 1.
    \label{eq:msedl}
\end{aligned}
\end{equation}

Then, we introduce $\rho_\mathrm{u}$ and $\rho_\mathrm{d}$ as the auxiliary variables, and obtain the transformed objective function shown in \eqref{eq:p3} on the basis of rate-WMMSE relationship.
\begin{subequations}
\label{eq:p3}
    \begin{align}
    &\underset{\substack{\{\mathbf{p}, \mathbf{w}\}, \{\boldsymbol{\omega}_\mathrm{u},\mathbf{u}_\mathrm{d}\},\\
    \{\rho_\mathrm{u},\rho_\mathrm{d}\}}}{\text{max}}
    && \alpha_1 (\log_2 \rho_\mathrm{u} - \rho_\mathrm{u} \mathsf{E}_\mathrm{BS}) + \alpha_2 (\log_2 \rho_\mathrm{d} - \rho_\mathrm{d} \mathsf{E}_\mathrm{d}) \nonumber\\
    &&& + \alpha_3 G_\mathrm{t} + \alpha_4 G_\mathrm{r} -\frac{1}{2\beta} \|\mathbf{w}^H \mathbf{H}_\mathrm{si} \mathbf{p} \|^2_2
    \label{eq:op3}\\
    &\qquad \text{s.t.}
    &&\mathrm{(\ref{eq:op1c1})-(\ref{eq:op1c3})}.\label{eq:op3c1} \nonumber
    \end{align}
\end{subequations}

\subsubsection{Beampattern Power Transformation}
\label{sssect:beam}
To make $G_\mathrm{t}$ and $G_\mathrm{r}$ concave, we reformulate them with $V_\mathrm{t} = \|\mathbf{b}(\theta_\mathrm{r})\|^2$ as 
\begin{equation}
\begin{aligned}
    G_\mathrm{t} &= |\mathbf{b}^H(\theta_\mathrm{r}) \mathbf{p}|^2 = \mathbf{p}^H \mathbf{b}(\theta_\mathrm{r})\mathbf{b}^H(\theta_\mathrm{r}) \mathbf{p}\\
    & = \mathbf{p}^H \mathbf{b}(\theta_\mathrm{r})\mathbf{b}^H(\theta_\mathrm{r}) \mathbf{p} - V_\mathrm{t} P_\mathrm{d} + V_\mathrm{t} P_\mathrm{d}\\
    & \leq \mathbf{p}^H \mathbf{b}(\theta_\mathrm{r})\mathbf{b}^H(\theta_\mathrm{r}) \mathbf{p} - \mathbf{p}^H V_\mathrm{t} \mathbf{I} \mathbf{p} + V_\mathrm{t} P_\mathrm{d}\\
    & = \mathbf{p}^H (\mathbf{b}(\theta_\mathrm{r}) \mathbf{b}^H(\theta_\mathrm{r})- V_\mathrm{t} \mathbf{I}) \mathbf{p} + V_\mathrm{t} P_\mathrm{d}\\
    & = \mathbf{p}^H Z_\mathrm{t}(\theta_\mathrm{r})\mathbf{p} + V_\mathrm{t} P_\mathrm{d},
\end{aligned}
\end{equation}
where we denote $Z_\mathrm{t}(\theta_\mathrm{r}) = \mathbf{b}(\theta_\mathrm{r}) \mathbf{b}^H(\theta_\mathrm{r})- V_\mathrm{t} \mathbf{I}$. The equality holds when the power of $\mathbf{p}$ reaches $P_\mathrm{d}$. With the expression of the steering vector, it is straightforward to know that $\mathbf{b}(\theta_\mathrm{r}) \mathbf{b}^H(\theta_\mathrm{r})$ is a rank-1 matrix with the eigenvalue $\|\mathbf{b}(\theta_\mathrm{r})\|^2$. Therefore, $Z_\mathrm{t}(\theta_\mathrm{r})$ is negative semi-definite, and $\mathbf{p}^H Z(\theta_\mathrm{r})\mathbf{p}$ is concave. Subsequently, maximizing $G_\mathrm{t}$ can be approximated as the maximization of $\mathbf{p}^H Z_\mathrm{t}(\theta_\mathrm{r})\mathbf{p}$ by ignoring the constant term $V_\mathrm{t} P_\mathrm{d}$.

Analogous to the reformulation of $G_\mathrm{t}$, the expression of $G_\mathrm{r}$ is given by
\begin{equation}
\begin{aligned}
    G_\mathrm{r} \leq \mathbf{w}^H Z_\mathrm{r}(\theta_\mathrm{r})\mathbf{w} + V_\mathrm{r},
\end{aligned}
\end{equation}
where $Z_\mathrm{r}(\theta_\mathrm{r})=\mathbf{a}(\theta_\mathrm{r}) \mathbf{a}^H_\mathrm{r}(\theta_\mathrm{r})- V_\mathrm{r} \mathbf{I} $ and $ V_\mathrm{r} = \|\mathbf{a}(\theta_\mathrm{r})\|^2$.  The equality holds when $\| \mathbf{w} \|^2_2 = 1$. Consequently, problem \eqref{eq:p1} is reformulated as problem \eqref{eq:p4} shown at the top of the next page.  

\begin{figure*}[!ht]
\begin{subequations}
\label{eq:p4}
    \begin{align}
    &\underset{\substack{\{\mathbf{p}, \mathbf{w}\}, \{\boldsymbol{\omega}_\mathrm{u},\mathbf{u}_\mathrm{d}\},\\
    \{\rho_\mathrm{u},\rho_\mathrm{d}\}}}{\text{max}}
    && \alpha_1 (\log_2 \rho_\mathrm{u} - \rho_\mathrm{u} \mathsf{E}_\mathrm{BS}) + \alpha_2 (\log_2 \rho_\mathrm{d} - \rho_\mathrm{d} \mathsf{E}_\mathrm{d}) + \alpha_3 \mathbf{p}^H Z_\mathrm{t}(\theta_\mathrm{t})\mathbf{p} + \alpha_4 \mathbf{w}^H Z_\mathrm{r}(\theta_\mathrm{r})\mathbf{w} -\frac{1}{2\beta} \|\mathbf{w}^H \mathbf{H}_\mathrm{si} \mathbf{p} \|^2_2 \label{eq:op4} \\
    &\qquad \text{s.t.}
    &&\eqref{eq:op1c1}-\eqref{eq:op1c3}. \nonumber
    \end{align}
    \hrulefill
\end{subequations}
\end{figure*}

\subsection{Penalty-based Block Update Algorithm}
\label{subsec:block}
With the BCD method, an iterative approach is utilized where each variable is optimized while keeping the others fixed, till convergence.

\subsubsection{Block $ \{ \rho_\mathrm{u}$, $\rho_\mathrm{d} \}$}
When $\{\mathbf{p}, \mathbf{w}\}$, $\{\boldsymbol{\omega}_\mathrm{u},\mathbf{u}_\mathrm{d}\}$ are fixed, the two separate sub-problems with regards to $\rho_\mathrm{u}$ and $\rho_\mathrm{d}$ are both convex and unconstrained given by
\begin{maxi}|s|                   
    {\rho_\mathrm{u}}                               
    {\log_2\rho_\mathrm{u} - \rho_\mathrm{u} \mathsf{E}_\mathrm{BS} \label{eq:bcdmseu}}   
    {}             
    {},
\end{maxi}
and
\begin{maxi}|s|                   
    {\rho_\mathrm{d}}                               
    {\log_2\rho_\mathrm{d} - \rho_\mathrm{d} \mathsf{E}_\mathrm{d} \label{eq:bcdmsed}}   
    {}             
    {}.
\end{maxi}
Thus, the optimal $\rho_\mathrm{u}^*$ and $\rho_\mathrm{d}^*$ can be obtained by setting the partial derivative with respect to the two optimized variables to zero. Subsequently, the expressions for $\rho_\mathrm{u}^*$ and $\rho_\mathrm{d}^*$ are:
\begin{equation}
\begin{aligned}
\qquad \quad    \rho_\mathrm{u}^* &= \mathsf{E}_\mathrm{BS}^{-1} \\
    & = (\mathbf{w}^H \mathbf{H}_\mathrm{u} \boldsymbol{\omega}_\mathrm{u} \boldsymbol{\omega}_\mathrm{u}^H \mathbf{H}_\mathrm{u}^H \mathbf{w} - 2 \Re\{\mathbf{w}^H \mathbf{H}_\mathrm{u} \boldsymbol{\omega}_\mathrm{u} \}\\
    & \quad + \mathbf{w}^H (\mathbf{H} \mathbf{p} \mathbf{p}^H \mathbf{H}^H + \sigma^2_\mathrm{u}\mathbf{I}) \mathbf{w} +1)^{-1},
\end{aligned}
\label{eq:rho_u}
\end{equation}
\begin{equation}
\begin{aligned}
    \rho_\mathrm{d}^* &= \mathsf{E}_\mathrm{d}^{-1} \\
    & = (\mathbf{u}^H_\mathrm{d} \mathbf{H}_\mathrm{d} \mathbf{p} \mathbf{p}^H \mathbf{H}^H_\mathrm{d} \mathbf{u}_\mathrm{d} - 2 \Re \{\mathbf{u}^H_\mathrm{d}  \mathbf{H}_\mathrm{d} \mathbf{p}\}\\
    & \quad + \sigma^2_\mathrm{d} \mathbf{u}^H_\mathrm{d} \mathbf{u}_\mathrm{d} + 1)^{-1}.
\end{aligned}
\label{eq:rho_d}
\end{equation}
The complexity of \eqref{eq:rho_u} and \eqref{eq:rho_d} are $\mathcal{O}(N_\mathrm{r}^2)$, and $\mathcal{O}(N_\mathrm{t}  N_\mathrm{d})$, respectively, mainly due to matrix multiplication.
\subsubsection{Block $ \{ \boldsymbol{\omega}_\mathrm{u} \}$}
\label{sssec:omega}
With fixed $\{ \mathbf{p}$, $\mathbf{w} \}$, $\{ \rho_\mathrm{u}$, $\rho_\mathrm{d} \}$, the sub-problem with regard to $\boldsymbol{\omega}_\mathrm{u}$ is given by
\begin{mini}|s|[2]                   
    {\boldsymbol{\omega}_\mathrm{u}}                               
    {\alpha_1 \mathsf{E}_\mathrm{BS} \label{eq:bcd1}}   
    {}             
    {} 
    \addConstraint{\| \boldsymbol{\omega}_\mathrm{u} \|^2_2}{\leq P_\mathrm{u}}.
\end{mini}
Since the objective function is convex with the convex constraint, the Lagrange multiplier method based on the  Karush–Kuhn–Tucker (KKT) condition is used with expression:
\begin{mini}|s|[2]                   
    {\boldsymbol{\omega}_\mathrm{u}}                               
    {\alpha_1 \mathsf{E}_\mathrm{BS} + \mu (\| \boldsymbol{\omega}_\mathrm{u} \|^2_2 - P_\mathrm{u}) \label{eq:lar2o}}   
    {}             
    {} .
\end{mini}
To solve this problem, we first substitute \eqref{eq:lar2o} with \eqref{eq:msebs}. Subsequently, we take the partial derivative of the Lagrangian function with regard to $\boldsymbol{\omega}_\mathrm{u}$ and $\mu$, respectively, and set both to zero. As a result, we obtain
\begin{equation}
    \boldsymbol{\omega}_\mathrm{u} = (\alpha_1 \mathbf{w}^H \mathbf{H}_\mathrm{u} \mathbf{H}_\mathrm{u}^H \mathbf{w} + \mu)^{-1} \alpha_1  \mathbf{H}_\mathrm{u}^H \mathbf{w},
    \label{eq:omega_u_1}
\end{equation}
and
\begin{equation}
    \| \boldsymbol{\omega}_\mathrm{u} \|^2_2 = P_\mathrm{u}.
    \label{eq:bilambda}
\end{equation}
By substituting \eqref{eq:bilambda} with \eqref{eq:omega_u_1}, we can then utilize the bisection search to obtain $\mu^*$. Thereon, the optimal solution $\boldsymbol{\omega}_\mathrm{u}^*$ is obtained by replacing $\mu$ with $\mu^*$ in \eqref{eq:omega_u_1}.
\begin{equation}
    \boldsymbol{\omega}_\mathrm{u}^* = (\alpha_1 \mathbf{w}^H \mathbf{H}_\mathrm{u} \mathbf{H}_\mathrm{u}^H \mathbf{w} + \mu^*)^{-1} \alpha_1 \mathbf{H}_\mathrm{u}^H \mathbf{w}.
    \label{eq:omega_u}
\end{equation}
Updating $\boldsymbol{\omega}_\mathrm{u}$ requires $\mathcal{O}(I_1 N_\mathrm{r} N_\mathrm{u})$ complexity due to the bisection search for $\mu$ with $I_1$ the number of iterations and the matrix multiplication.

\subsubsection{Block $\{ \mathbf{w} \}$}
Under the condition that $\mathbf{p}$, $\boldsymbol{\omega}_\mathrm{u}$, $\mathbf{u}_\mathrm{d}$, $\rho_\mathrm{u}$, and $\rho_\mathrm{d}$ are all fixed, $\mathbf{w}$ is only related to the first and the last two terms in \eqref{eq:op4}, yielding the following convex sub-problem
\begin{mini}|s|[2]                   
    {\mathbf{w}}                               
    {\alpha_1 \rho_\mathrm{u} \mathsf{E}_\mathrm{BS} + \alpha_4 \mathbf{w}^H Z_\mathrm{r}(\theta_\mathrm{r}) \alpha_4 \mathbf{w} + \frac{1}{2\beta} \|\mathbf{w}^H \mathbf{H}_\mathrm{si} \mathbf{p} \|^2_2 \label{eq:bcd3o}}   
    {}             
    {} .
\end{mini}
Although the constraint \eqref{eq:op1c3} that $\|\mathbf{w}\|^2_2  =1$ is a non-convex constraint, we firstly address this sub-problem with respect to $\mathbf{w}$ as an unconstrained complex optimization problem, and normalize the obtained solution after the convergence of the algorithm is satisfied.
Equating the partial derivative of \eqref{eq:bcd3o} with respect to $\mathbf{w}$ to zero, the optimal $\mathbf{w}^*$ is given by
\begin{equation}
    \mathbf{w}^* = (2 \mathbf{X}_\mathrm{u} + \frac{1}{\beta} \mathbf{H}_\mathrm{si} \mathbf{p} \mathbf{p}^H \mathbf{H}_\mathrm{si}^H)^{-1} 
    2 \alpha_1 \rho_\mathrm{u} \mathbf{H}_\mathrm{u} \boldsymbol{\omega}_\mathrm{u},
    \label{eq:w}
\end{equation}
where $\mathbf{X}_\mathrm{u}$ is defined as 
\begin{equation}
    \begin{aligned}
        \mathbf{X}_\mathrm{u} & = \alpha_1 \rho_\mathrm{u} \mathbf{H}_\mathrm{u} \boldsymbol{\omega}_\mathrm{u} \boldsymbol{\omega}_\mathrm{u}^H \mathbf{H}_\mathrm{u}^H
        + \alpha_1 \rho_\mathrm{u} \mathbf{H} \mathbf{p} \mathbf{p}^H \mathbf{H}^H \\
        & \quad + \alpha_1 \rho_\mathrm{u} \sigma_\mathrm{u}^2 \mathbf{I} - \alpha_4 Z_\mathrm{r}(\theta_\mathrm{r}).
    \end{aligned}
\end{equation}
The complexity of this step is $\mathcal{O}(N_\mathrm{r}^3)$ mainly due to the matrix inversion.

\subsubsection{Block $\{ \mathbf{u}_\mathrm{d} \}$} 
The uplink precoder is only related to the second term in \eqref{eq:op4} with other blocks fixed. Therefore, the unconstrained convex sub-problem is given by
\begin{mini}|s|[2]    
{\mathbf{u}_\mathrm{d}}{\mathbf{u}^H_\mathrm{d} \mathbf{H}_\mathrm{d} \mathbf{p} \mathbf{p}^H \mathbf{H}^H_\mathrm{d} \mathbf{u}_\mathrm{d} - 2 \Re \{\mathbf{u}^H_\mathrm{d}  \mathbf{H}_\mathrm{d} \mathbf{p}\}}{}{}
\breakObjective{+ \sigma^2_\mathrm{d} \mathbf{u}^H_\mathrm{d} \mathbf{u}_\mathrm{d} + 1.}
{\label{eq:bcd4o}}
\end{mini}
The optimal solution is given by 
\begin{equation}
    \mathbf{u}_\mathrm{d}^* = ( \mathbf{H}_\mathrm{d}  \mathbf{p}  \mathbf{p}^H \mathbf{H}_\mathrm{d}^H + \sigma^2_\mathrm{d} \mathbf{I})^{-1} \mathbf{H}_\mathrm{d} \mathbf{p}.
    \label{eq:u_d}
\end{equation}
Updating the $\mathbf{u}_\mathrm{d}$ in \eqref{eq:u_d} requires $\mathcal{O}(N_\mathrm{t} N_\mathrm{d})$ complexity mainly due to matrix multiplication.

\subsubsection{Block $\{ \mathbf{p} \}$} 
Under fixed $\rho_\mathrm{u}$, $\rho_\mathrm{d}$, $\mathbf{w}$, $\boldsymbol{\omega}_\mathrm{u}$ and $\mathbf{u}_\mathrm{d}$, the sub-problem with respect to $\mathbf{p}$ is a convex optimization problem with a convex constraint on the transceiver transmit power, $P_\mathrm{d}$, as follows
\begin{mini}|s|[2]    
{\mathbf{p}}{\alpha_1 \rho_\mathrm{u} \mathsf{E}_\mathrm{BS} + \alpha_2 \rho_\mathrm{d} \mathsf{E}_\mathrm{d} - \alpha_3 \mathbf{p}^H Z_\mathrm{t}(\theta_\mathrm{t}) \mathbf{p} + \frac{1}{2 \beta} \| \mathbf{w}^H \mathbf{H}_\mathrm{si} \mathbf{p} \|^2}{}{}
\addConstraint{ \| \mathbf{p} \|^2_2}{\leq P_\mathrm{d}}{\label{eq:bcd5}}.
\end{mini}

Similar to the design of block $\{ \boldsymbol{\omega}_\mathrm{u}  \}$ in \ref{sssec:omega}, this sub-problem with respect to $\mathbf{p}$ can be reformulated by the Lagrange multiplier method based on the KKT condition written below:
\begin{mini}|s|[2]    
{\mathbf{p}}{\alpha_1 \rho_\mathrm{u} \mathsf{E}_\mathrm{BS} + \alpha_2 \rho_\mathrm{d} \mathsf{E}_\mathrm{d} - \alpha_3 \mathbf{p}^H Z_\mathrm{t}(\theta_\mathrm{t}) \mathbf{p} + \frac{1}{2 \beta} \| \mathbf{w}^H \mathbf{H}_\mathrm{si} \mathbf{p} \|^2}{}{}
\breakObjective{+ \Gamma ( \| \mathbf{p} \|^2_2 - P_\mathrm{d}).}
{\label{eq:lar5}}
\end{mini}
By taking the partial derivative with respect to $\Gamma$ and $\mathbf{p}$, respectively, and setting them to zero, we obtain
\begin{equation}
\label{eq:p}
     \| \mathbf{p} \|^2_2 = P_\mathrm{d},
\end{equation}
and
\begin{equation}
    \begin{aligned}
        \mathbf{p} &= (2 \mathbf{X}_\mathrm{d} + \frac{1}{\beta}(\mathbf{H}_\mathrm{si}^H \mathbf{w} \mathbf{w}^H \mathbf{H}_\mathrm{si}) + 2 \Gamma \mathbf{I})^{-1}\\
        & \quad \cdot (2 \alpha_2 \rho_\mathrm{d} \mathbf{H}_\mathrm{d}^H \mathbf{u}_\mathrm{d}),
    \end{aligned}
    \label{eq:p_1}
\end{equation}
where $\mathbf{X}_\mathrm{d}$ is defined in \eqref{eq:xd} for expression simplicity,
\begin{equation}
    \mathbf{X}_\mathrm{d} = \alpha_1 \rho_\mathrm{u} \mathbf{H}^H \mathbf{w} \mathbf{w}^H \mathbf{H} + \alpha_2 \rho_\mathrm{d} \mathbf{H}_\mathrm{d} \mathbf{u}_\mathrm{d} \mathbf{u}_\mathrm{d}^H \mathbf{H}_\mathrm{d}^H
    - \alpha_3 Z_\mathrm{t}(\theta_\mathrm{t}) \mathbf{I}.
    \label{eq:xd}
\end{equation}
Subsequently, we substitute \eqref{eq:p_1} into \eqref{eq:p}, and utilize the bisection search to obtain the optimal $\Gamma^*$. Finally, the optimal $\mathbf{p}^*$ is given by
\begin{equation}
    \begin{aligned}
        \mathbf{p^*} &= (2 \mathbf{X}_\mathrm{d} + \frac{1}{\beta}(\mathbf{H}_\mathrm{si}^H \mathbf{w} \mathbf{w}^H \mathbf{H}_\mathrm{si}) + 2 \Gamma^* \mathbf{I})^{-1}\\
        & \quad \cdot (2 \alpha_2 \rho_\mathrm{d} \mathbf{H}_\mathrm{d}^H \mathbf{u}_\mathrm{d}).
    \end{aligned}
    \label{eq:p*}
\end{equation}
The complexity of updating $\mathbf{p}$ in \eqref{eq:p*} is $\mathcal{O}(I_2 N_\mathrm{t}^3)$ caused by the matrix inversion, where $I_2$ denotes the number of iterations in the bisection search for $\Gamma$.
\subsubsection{Summary}
Following the framework of penalty-based approach, iterative BCD update process, and setting appropriate initial value of the optimized variables, $\rho_\mathrm{u}$, $\rho_\mathrm{d}$, $\mathbf{w}$, $\boldsymbol{\omega}_\mathrm{u}$, $\mathbf{u}_\mathrm{d}$, and $\mathbf{p}$, the optimal solution in each block can be found after reaching the convergence condition. To satisfy the receive power constraint of $\mathbf{w}$, we normalize $\mathbf{w}$ after the whole iteration as below:
\begin{equation}
    \mathbf{w} = \frac{\mathbf{w}}{\| \mathbf{w} \|}.
    \label{eq:w_nor}
\end{equation}

\begin{algorithm}[t]
	\caption{Proposed Penalty-based Joint Transmit and Receive Beamformer Design}
	\label{alg:alg1}
	\KwIn{ $\mathbf{H}_\mathrm{d}$, $\mathbf{H}_\mathrm{u}$, $\mathbf{H}_\mathrm{r}$, $\mathbf{H}_\mathrm{si}$, $P_\mathrm{d}$, $P_\mathrm{u}$.}  
	\KwOut{$\rho_\mathrm{u}^*, \rho_\mathrm{d}^*, \mathbf{p}^*, \mathbf{w}^*, \boldsymbol{\omega}_\mathrm{u}^*, \mathbf{u}_\mathrm{d}^*$.} 
	\BlankLine
	Initialize $ \mathbf{p}, \mathbf{w}, \boldsymbol{\omega}_\mathrm{u}, \mathbf{u}_\mathrm{d}$ randomly, $\beta=10^{-25}$;
	
	\While{\textnormal{no convergence of objective function \eqref{eq:op4}}}{
        Update $ \rho_\mathrm{u}^*$ by \eqref{eq:rho_u}. Complexity = $\mathcal{O}(N_\mathrm{r}^2)$. \\
        Update $ \rho_\mathrm{d}^*$ by \eqref{eq:rho_d}. Complexity = $\mathcal{O}(N_\mathrm{t}  N_\mathrm{d})$.\\
		Update $\boldsymbol{\omega}_\mathrm{u}^*$ by \eqref{eq:omega_u}. Complexity = $\mathcal{O}(I_1 N_\mathrm{r} N_\mathrm{u})$.\\
		Update $\mathbf{w}^*$ by \eqref{eq:w}. Complexity = $\mathcal{O}(N_\mathrm{r}^3)$.\\
		Update $\mathbf{u}_\mathrm{d}^*$ by \eqref{eq:u_d}. Complexity = $\mathcal{O}(N_\mathrm{t} N_\mathrm{d}).$\\
		Update $\mathbf{p}^*$ by \eqref{eq:p*}. Complexity = $\mathcal{O}(I_2 N_\mathrm{t}^3)$.
	}
	
	Normalize $\mathbf{w}^*$ by \eqref{eq:w_nor}; 
	
	Return $\rho_\mathrm{u}^*, \rho_\mathrm{d}^*, \mathbf{p}^*, \mathbf{w}^*, \boldsymbol{\omega}_\mathrm{u}^*, \mathbf{u}_\mathrm{d}^*$.
	
\end{algorithm}

We summarize our proposed joint ISAC TX-RX beamformers design algorithm in Algorithm \ref{alg:alg1}.

\begin{remark}
An alternative to \eqref{eq:op1c4} is to restrict the residual SI power to below a threshold (i.s., $\frac{1}{2\beta} \|\mathbf{w}^H \mathbf{H}_\mathrm{si} \mathbf{p} \|^2_2 < \varepsilon$), which has been shown to improve communications performance in \cite{taghizadeh2016transmit}. However, we have adopted a stronger constraint in \eqref{eq:op1c4} because with the inclusion of sensing, where the radar returns are both very weak and strongly correlated with the residual SI, the latter needs to be suppressed to a far greater extent.

Nevertheless, enforcing such a strong constraint deteriorates both communications and sensing performance. Our compromise to this trade-off is detailed in the following comment.
\end{remark}

\begin{remark} \textbf{Comparison between PDD and Algorithm \ref{alg:alg1}}
PDD is a widely-used technique to solve optimization problems involving non-smooth, non-convex functions \cite{shi2020penalty}. Algorithm \ref{alg:alg1} is similar to PDD due to the same problem formulation as shown in \eqref{eq:op2}. However, the PDD is a two-loop iterative algorithm, wherein the inner loop solves an augmented Lagrangian problem by BCD method (i.e., the steps within the while loop in Algorithm \ref{alg:alg1}), and the outer loop updates the dual variable and penalty term until convergence(i.e., $\lambda \mathbf{w}^H \mathbf{H}_\mathrm{si} \mathbf{p} + \frac{1}{2\beta} \|\mathbf{w}^H \mathbf{H}_\mathrm{si} \mathbf{p} \|^2_2 < \varepsilon$). In Algorithm \ref{alg:alg1}, we omit the outer loop iteration of the PDD framework because for sufficiently small $\beta$, the residual SI power can be suppressed below a suitable threshold (e.g. noise floor). In contrast, having the outer loop  would needlessly suppress the residual SI further (i.e., enforcing \eqref{eq:op1c4}), which not only reduces both communications and sensing performance, but also leads to slower convergence. 
\end{remark}

\begin{remark}
In general, the ISAC TX-RX beamformer design problem can be formulated in several ways; for instance, an alternative to \eqref{eq:op1}-\eqref{eq:op1c4} is to maximize communications performance to constraints on radar sensing performance. 

However, an advantage of our formulation is that it permits a closed-form solution for each step of the BCD (i.e., each of the subproblems in Section \ref{subsec:block}). On the other hand, incorporating the radar sensing performance as a constraint would have made it difficult to have realize this, and thus slowed down the convergence of our Algorithm. On the other hand, a softer constraint on sensing performance can be imposed by increasing the weights $\alpha_3$ and $\alpha_4$, which would have the effect of prioritizing the sensing performance.
\end{remark}

\begin{remark}
The joint TX-RX beamformer design can be extended to underloaded multi-user scenarios as well. For example, consider K downlink users and let $\mathbf{p}_k$ denote the downlink precoder for the $k^{\text{th}}$ downlink user and $\mathbf{P} := [\mathbf{p}_1, \cdots \mathbf{p}_K] \in \mathbb{C}^{N_t \times K}$. For $K < \min (N_t, N_r)$, the rows of $\mathbf{H}_\mathrm{si} \mathbf{P}$ have a non-trivial nullspace and hence, there exists non-zero $\mathbf{w}$ for which $\mathbf{w}^H \mathbf{H}_\mathrm{si} \mathbf{P}$ = 0. 
\end{remark}



\subsection{Algorithm Implementation}
In this paper, we consider a centralized implementation of Algorithm \ref{alg:alg1}, which involves the overhead of communicating all the channel state information ($\mathbf{H}_\mathrm{u}$, $\mathbf{H}_\mathrm{si}$, $\mathbf{H}_d$) and the target direction to a controller as well as transferring the optimized $\mathbf{p}$ and $\mathbf{w}$ to the ISAC transceiver, $\boldsymbol{\omega}_{\mathrm u}$ to the uplink user and $\boldsymbol{\omega}_{\mathrm d}$ to the downlink user. Thus, a decentralized implementation of Algorithm \ref{alg:alg1} is key for minimizing the overhead. Furthermore, the order in which the variables are optimized can be different, and thus, a different ordering along with possibly different update rates may speed up convergence. These questions are left for future work. In the following section, we discuss the convergence behavior and the complexity of Algorithm \ref{alg:alg1}.


\section{Convergence and Complexity Analysis}
\label{sec:analysis}
\subsection{Convergence Analysis}
In this section, the convergence of Algorithm \ref{alg:alg1} to at least a local optimum is proved. Let $f(\mathbf{p}, \mathbf{w}, \boldsymbol{\omega}_\mathrm{u}, \mathbf{u}_\mathrm{d})$ denote the objective function in \eqref{eq:op2}, and $f_\rho(\mathbf{p}, \mathbf{w}, \boldsymbol{\omega}_\mathrm{u}, \mathbf{u}_\mathrm{d}, \rho_\mathrm{u}, \rho_\mathrm{d})$ the reformulated objective function in \eqref{eq:op3}, based on the rate-WMMSE relationship explained in Section \ref{sssec:wmmse}. Thus, we have
\begin{equation}
\label{eq:ineq1}
\begin{aligned}
f(\mathbf{p}, \mathbf{w}, \boldsymbol{\omega}_\mathrm{u}, \mathbf{u}_\mathrm{d}) 
& \geq f_\rho(\mathbf{p}, \mathbf{w}, \boldsymbol{\omega}_\mathrm{u}, \mathbf{u}_\mathrm{d}, \rho_\mathrm{u}, \rho_\mathrm{d}),
\end{aligned}
\end{equation}
where $\rho_\mathrm{u}^*$, and $\rho_\mathrm{d}^*$ are the optimal solutions as shown in \eqref{eq:rho_u} and \eqref{eq:rho_d}. The inequality condition in \eqref{eq:ineq1} holds when $\rho_\mathrm{u}$ satisfies \eqref{eq:rho_u} and $\rho_\mathrm{d}$ satisfies \eqref{eq:rho_d} are both the global optimum with fixed other optimized variables $\mathbf{p}, \mathbf{w}, \boldsymbol{\omega}_\mathrm{u}, \mathbf{u}_\mathrm{d}$. Let $f_\mathrm{g}(\mathbf{p}, \mathbf{w}, \boldsymbol{\omega}_\mathrm{u}, \mathbf{u}_\mathrm{d}, \rho_\mathrm{u}, \rho_\mathrm{d})$ denote the objective in \eqref{eq:op4}. Then we have
\begin{equation}
    \begin{aligned}
    f_\rho(\mathbf{p}, \mathbf{w}, \boldsymbol{\omega}_\mathrm{u}, \mathbf{u}_\mathrm{d}, \rho_\mathrm{u}, \rho_\mathrm{d}) & \le
    f_\mathrm{g}(\mathbf{p}, \mathbf{w}, \boldsymbol{\omega}_\mathrm{u}, \mathbf{u}_\mathrm{d}, \rho_\mathrm{u}, \rho_\mathrm{d}) \\
    & \quad + \alpha_3 V_\mathrm{t} P_\mathrm{d} + \alpha_4 V_\mathrm{r}.
    \end{aligned}
    \label{eq:equ1}
\end{equation}

Let $\mathbf{p}^{(\mathrm{n})}, \mathbf{w}^{(\mathrm{n})}, \boldsymbol{\omega}_\mathrm{u}^{(\mathrm{n})}, \mathbf{u}_\mathrm{d}^{(\mathrm{n})}, \rho_\mathrm{u}^{(\mathrm{n})}, \rho_\mathrm{d}^{(\mathrm{n})}$ be the n-th iteration of the variables whose expressions are given by \eqref{eq:p*}, \eqref{eq:w}, \eqref{eq:omega_u}, \eqref{eq:u_d}, \eqref{eq:rho_u} and \eqref{eq:rho_d}, respectively.  $\rho_\mathrm{u}^{(\mathrm{n})}, \rho_\mathrm{d}^{(\mathrm{n})}$ are obtained by the n-th iteration beamformers $\{\mathbf{p}^{(\mathrm{n})}, \mathbf{w}^{(\mathrm{n})}, \boldsymbol{\omega}_\mathrm{u}^{(\mathrm{n})} \}$ and $\{\mathbf{p}^{(\mathrm{n})},  \mathbf{u}_\mathrm{d}^{(\mathrm{n})} \}$, respectively. Since the power of $\mathbf{p}^{\mathrm{(n)}}$ reaches $P_\mathrm{d}$, we obtain \eqref{eq:f_fg} from \eqref{eq:ineq1} and \eqref{eq:equ1}, given by
\begin{equation}
\label{eq:f_fg}
    \begin{aligned}
        &f(\mathbf{p}^{(\mathrm{n+1})}, \mathbf{w}^{(\mathrm{n+1})}, \boldsymbol{\omega}_\mathrm{u}^{(\mathrm{n+1})}, \mathbf{u}_\mathrm{d}^{(\mathrm{n+1})}) \\
        & \geq f_\mathrm{g}(\mathbf{p}^{(\mathrm{n+1})}, \mathbf{w}^{(\mathrm{n+1})}, \boldsymbol{\omega}_\mathrm{u}^{(\mathrm{n+1})}, \mathbf{u}_\mathrm{d}^{(\mathrm{n+1})}, \rho_\mathrm{u}^{(\mathrm{n})}, \rho_\mathrm{d}^{(\mathrm{n})}) \\
        & \quad + \alpha_3 V_\mathrm{t} P_\mathrm{d} + \alpha_4 V_\mathrm{r}.
    \end{aligned}
\end{equation}

Following Algorithm \ref{alg:alg1}, the update sequences are $\boldsymbol{\omega}_\mathrm{u}^{(\mathrm{n+1})} \overset{\text{(a)}} \leftarrow \{\mathbf{w}^{(\mathrm{n})} \}$, 
$\mathbf{w}^{(\mathrm{n+1})} \overset{\text{(b)}} \leftarrow \{\mathbf{p}^{(\mathrm{n})},  \boldsymbol{\omega}_\mathrm{u}^{(\mathrm{n})}, \rho_\mathrm{u}^{(\mathrm{n})}\}$,
$\mathbf{u}_\mathrm{d}^{(\mathrm{n+1})} \overset{\text{(c)}} \leftarrow \{\mathbf{p}^{(\mathrm{n})}\}$, and $\mathbf{p}^{(\mathrm{n+1})} \overset{\text{(d)}} \leftarrow \{ \rho_\mathrm{u}^{(\mathrm{n})}, \rho_\mathrm{d}^{(\mathrm{n})}, \mathbf{w}^{(\mathrm{n+1})}, \mathbf{u}_\mathrm{d}^{(\mathrm{n+1})} \}$. Therefore, we have
\begin{equation}
    \begin{aligned}
        &  f_\mathrm{g}(\mathbf{p}^{(\mathrm{n+1})}, \mathbf{w}^{(\mathrm{n+1})}, \boldsymbol{\omega}_\mathrm{u}^{(\mathrm{n+1})}, \mathbf{u}_\mathrm{d}^{(\mathrm{n+1})}, \rho_\mathrm{u}^{(\mathrm{n})}, \rho_\mathrm{d}^{(\mathrm{n})})\\
        & \geq f_\mathrm{g}(\mathbf{p}^{(\mathrm{n})}, \mathbf{w}^{(\mathrm{n})}, \boldsymbol{\omega}_\mathrm{u}^{(\mathrm{n})}, \mathbf{u}_\mathrm{d}^{(\mathrm{n})}, \rho_\mathrm{u}^{(\mathrm{n})}, \rho_\mathrm{d}^{(\mathrm{n})}).
    \end{aligned}
    \label{eq:itern}
\end{equation}
Subsequently, by using \eqref{eq:ineq1} and \eqref{eq:equ1} and due to the power of $\mathbf{p}^{\mathrm{(n)}}$ reaching $P_\mathrm{d}$, we have
\begin{equation}
\label{eq:fgf}
    \begin{aligned}
        & f_\mathrm{g}(\mathbf{p}^{(\mathrm{n})}, \mathbf{w}^{(\mathrm{n})}, \boldsymbol{\omega}_\mathrm{u}^{(\mathrm{n})}, \mathbf{u}_\mathrm{d}^{(\mathrm{n})}, \rho_\mathrm{u}^{(\mathrm{n})}, \rho_\mathrm{d}^{(\mathrm{n})}) + \alpha_3 V_\mathrm{t}  P_\mathrm{d} + \alpha_4 V_\mathrm{r}\\
        & = f(\mathbf{p}^{(\mathrm{n})}, \mathbf{w}^{(\mathrm{n})}, \boldsymbol{\omega}_\mathrm{u}^{(\mathrm{n})}, \mathbf{u}_\mathrm{d}^{(\mathrm{n})}).
    \end{aligned}
\end{equation}
From \eqref{eq:ineq1} - \eqref{eq:fgf}, we obtain
\begin{equation}
    f(\mathbf{p}^{(\mathrm{n+1})}, \mathbf{w}^{(\mathrm{n+1})}, \boldsymbol{\omega}_\mathrm{u}^{(\mathrm{n+1})}, \mathbf{u}_\mathrm{d}^{(\mathrm{n+1})}) \geq f(\mathbf{p}^{(\mathrm{n})}, \mathbf{w}^{(\mathrm{n})}, \boldsymbol{\omega}_\mathrm{u}^{(\mathrm{n})}, \mathbf{u}_\mathrm{d}^{(\mathrm{n})}),
\end{equation}
which shows that the objective function in \eqref{eq:op2} is non-decreasing after each BCD update iteration. Since the uplink and downlink rates are upper bounded \cite{clerckx2013mimo}, as are the beampattern gains when beamformers $\mathbf{p}$ and $\mathbf{w}$ both probe at target direction, $\theta_\mathrm{r}$, the value of objective in \eqref{eq:op2} is bounded above. Consequently, the proposed algorithm will converge to at least a local optimum, which guarantees the convergence of the proposed Algorithm \ref{alg:alg1}. In the following, we provide numerical results to demonstrate the convergence of the proposed design.

\begin{figure}[tb]
	\begin{center}
		\includegraphics[width =0.45\textwidth]{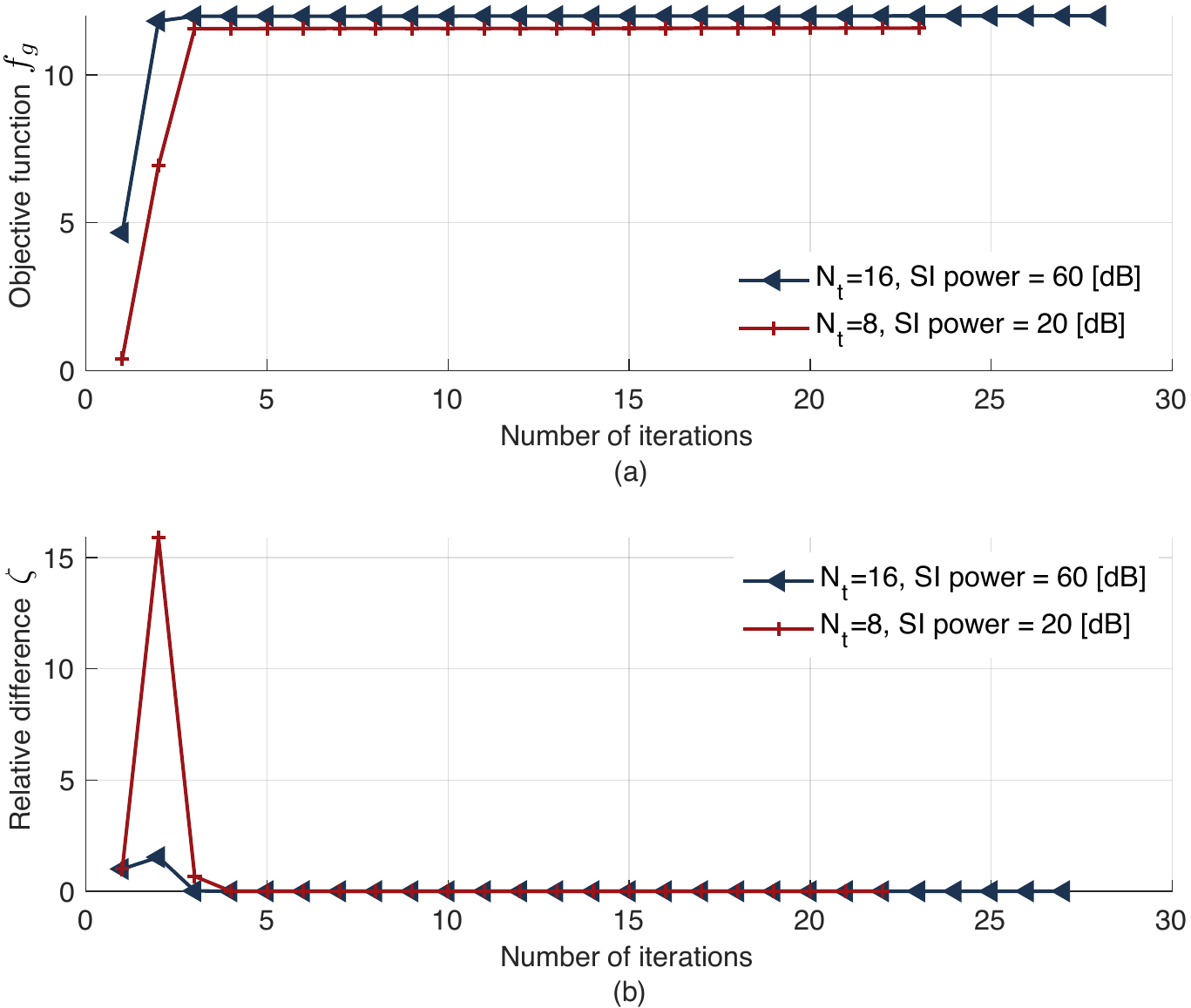}
		\caption{Convergence of Algorithm \ref{alg:alg1} with $\epsilon = 10^{-5}$. (a) Objective function versus the number of iterations; (b) Relative difference versus the number of iterations.}
		\label{fig:converge}
	\end{center}
\end{figure}

Fig. \ref{fig:converge} shows the behavior of the objective function \eqref{eq:op2} and the relative difference $\zeta$, defined in \eqref{eq:zeta}, with respect to the number of iterations, for different $N_\mathrm{t}$ and SI power levels. 
\begin{equation}
    \zeta \triangleq \frac{f_\mathrm{g}^\mathrm{(n)} - f_\mathrm{g}^\mathrm{(n-1)}}{f_\mathrm{g}^\mathrm{(n-1)}},
    \label{eq:zeta}
\end{equation}
where $\zeta \leq \epsilon$ is the iteration end criterion. As shown in Fig. \ref{fig:converge}, objective \eqref{eq:op4} converges within a small number of iterations.

\subsection{Complexity Analysis}
From Algorithm \ref{alg:alg1}, the bottleneck step is the step 8, which updates $\mathbf{p}$, with assumption $N_\mathrm{d} < N_\mathrm{t} \leq N_\mathrm{r}$, $N_\mathrm{u} < N_\mathrm{t} \leq N_\mathrm{r}$. Thus, the overall complexity of Algorithm \ref{alg:alg1} is given by
\begin{equation}
    \mathcal{O} (I_3 (N_\mathrm{r}^2 + N_\mathrm{t}  N_\mathrm{d} + I_1 N_\mathrm{r} N_\mathrm{u} + N_\mathrm{r}^3 + I_2 N_\mathrm{t}^3 ) )
    {\approx} \mathcal{O} (I_3 ( I_2 N_\mathrm{t}^3 ) ),
\end{equation}
where $I_3$ is the iteration number. In summary, Algorithm \ref{alg:alg1} has polynomial complexity.


\section{Numerical evaluation}
\label{sec:result}
In this section, we provide numerical results to validate the performance of the proposed joint TX-RX beamformer design for the FD ISAC system. The number of transmit and receive antennas at the FD ISAC transceiver are set the same, which are $N_\mathrm{t}=N_\mathrm{r}=16$. The number of antennas at uplink and downlink user are set to $N_\mathrm{u}=2$ and $N_\mathrm{d}=2$, respectively. The carrier frequency of the FD ISAC system is $f_\mathrm{c} = 2.4$GHz. The sampling rate is $T_\mathrm{s}=1/\Omega$, where $\Omega=20$MHz is the bandwidth. We use the tuple $(\theta, r, v)$ to represent a target/user's coordinates and bearing w.r.t the transceiver, where $\theta$ denotes the direction, $r$ the range and $v$ the velocity. The radar target is assumed to be at $(45^\circ, 7.5\text{m}, 20\text{m/s})$, the uplink user at $(-50^\circ, 10\text{m}, 0\text{m/s})$, and the downlink user at  $(\theta_\mathrm{d}=-30^\circ, 100\text{m}, 0\text{m/s})$.
The transmit powers of the ISAC transceiver and uplink user are $P_\mathrm{d,[dBm]} = 20$dBm, and $P_\mathrm{u,[dBm]} = 10$dBm, respectively. The thermal noise floor at the ISAC receiver is $P_\mathrm{noise,[dBm]} = -94$dBm. The path loss (in dB) is modelled as follows
$\eta(d) = -20 \log_{10} (\lambda/(4 \pi d_0)) +10 n \log_{10} ({d}/{d_{0}})$, where $d_{0}=1$m is the reference distance, and $n=2.2$ is the path loss exponent \cite{xu2002spatial}, respectively. The SI power, $P_\mathrm{si,[dBm]} - P_\mathrm{noise,[dBm]}$, ranges from $10$ - $60$dB at the ISAC receiver. We run $1000$ times Monte Carlo simulations  with perfect channel state information (CSI) for observing the performances of the SI cancellation, SoI over SI, and sum-rates. The signals, $s_\mathrm{u}$, and $s_\mathrm{d}$, comprise uncoded i.i.d zero mean, unit-energy QPSK symbols.


\subsection{SI Power After Cancellation}
\label{sssection:si}

\begin{figure}[tb]
	\begin{center}
		\includegraphics[width =0.45\textwidth]{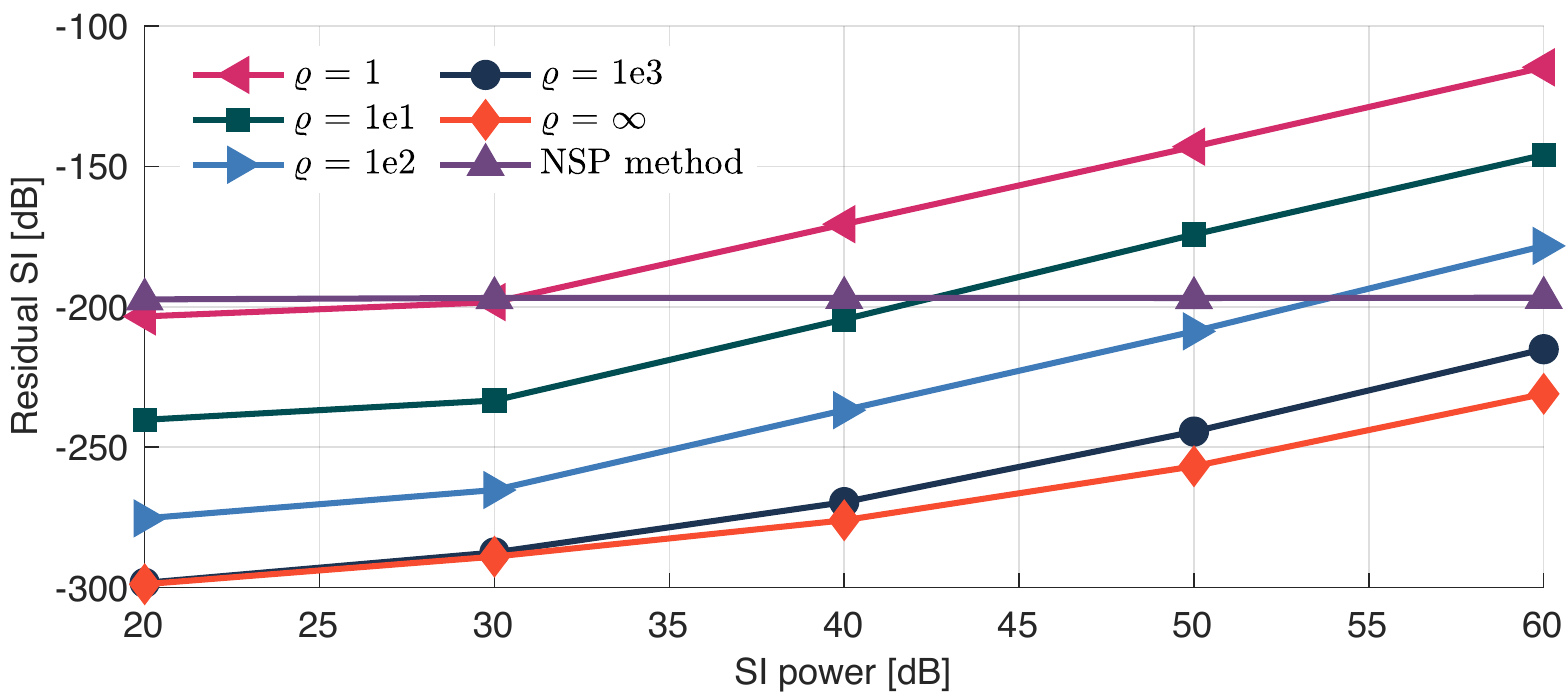}
		\caption{Performance comparison of SI cancellation with varying priority parameter $\varrho$ and level of SI. The residual SI level is computed with respect to the noise floor.}
		\label{fig:si}
	\end{center}
\end{figure}
\begin{figure}[tb]
	\begin{center}
		\includegraphics[width =0.45\textwidth]{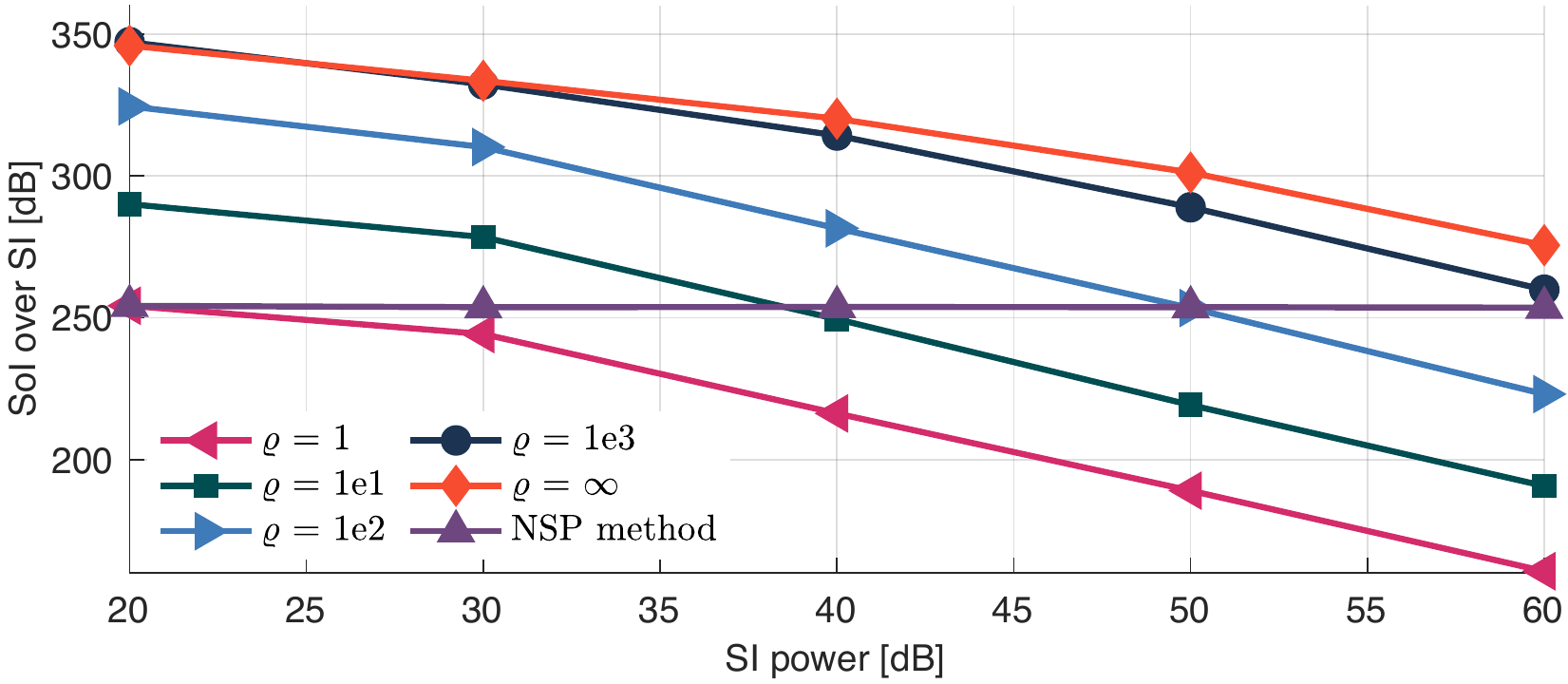}
		\caption{Performance comparison of SoI (e.g., radar echo and uplink data) power over SI power in dB.}
		\label{fig:soi}
	\end{center}
\end{figure}
Firstly, we measure the SI cancellation performance of Algorithm \ref{alg:alg1} in terms of the residual digital-domain SI power level $P_\mathrm{res,[dB]}$, which is defined as $P_\mathrm{res,[dB]} := 10\log_{10}(|\mathbf{w}^H \mathbf{H}_\mathrm{si} \mathbf{p}|^2 / 1\mathrm{mW}) - P_\mathrm{noise,[dBm]}$. Thus, a negative value of $P_\mathrm{res}$ implies that the residual SI power level is lower than the noise floor.  Additionally, we assume $\alpha_\mathrm{com} = \alpha_1 = \alpha_2$ (i.e., equal weightage for uplink and downlink communications performance) and $\alpha_\mathrm{radar} = \alpha_3 = \alpha_4$. Furthermore, the priority given to communications performance is captured by the parameter, $\varrho={\alpha_\mathrm{com}}/{\alpha_\mathrm{radar}}$. As shown in Fig. \ref{fig:si}, when the $\varrho$ varies from $1$ to $1000$, the SI power up to $60$ dB with regards to noise floor can be effectively suppressed due to the residual SI power, $P_\mathrm{res,[dB]} < 0$. To observe whether the SoI is preserved in the process of SI suppression, we consider the ratio between the SoI and SI powers, defined as $P_\mathrm{SoI/SI,[dB]} := 10\log_{10}( (|\mathbf{w}^H \mathbf{H}_\mathrm{r} \mathbf{p}|^2 + |\mathbf{w}^H \mathbf{H}_\mathrm{u} \boldsymbol{\omega}_\mathrm{u} |^2 ) / |\mathbf{w}^H \mathbf{H}_\mathrm{si} \mathbf{p}|^2 ) $. The performance of $P_\mathrm{SoI/SI,[dB]}$ with varying $\varrho$ is shown in Fig. \ref{fig:soi}, where a positive value implies that the SoI (e.g., radar echo and uplink data) is not drowned by the strong SI. As expected, the performance of SI cancellation degrades and $P_\mathrm{SoI/SI,[dB]}$ decreases with increasing of the SI power. Additionally, when the priority parameter $\varrho$ increases, the performance of SI cancellation improves and SoI over SI power increases. This is because a higher $\varrho$ leads to more priority given to communications, resulting in reduced correlation between SI and SoI. It should be noted that this correlation is one of the challenges in suppressing SI when the sensing function is present. Thus, the joint ISAC TX-RX beamformers design can effectively preserve SoI and suppress SI in FD ISAC. While NSP method has better performances than the proposed method with some certain parameter settings, other system capabilities of the proposed method (i.e., average sum-rate of downlink and uplink users, radar parameter estimation performance) are enhanced, which are explained in Section \ref{ssec:sumrate} and Section \ref{ssec:radaresti}, respectively. 

\subsection{Beampattern Power Performance}


\begin{figure*}
	\begin{center}
		\includegraphics[width =1\textwidth]{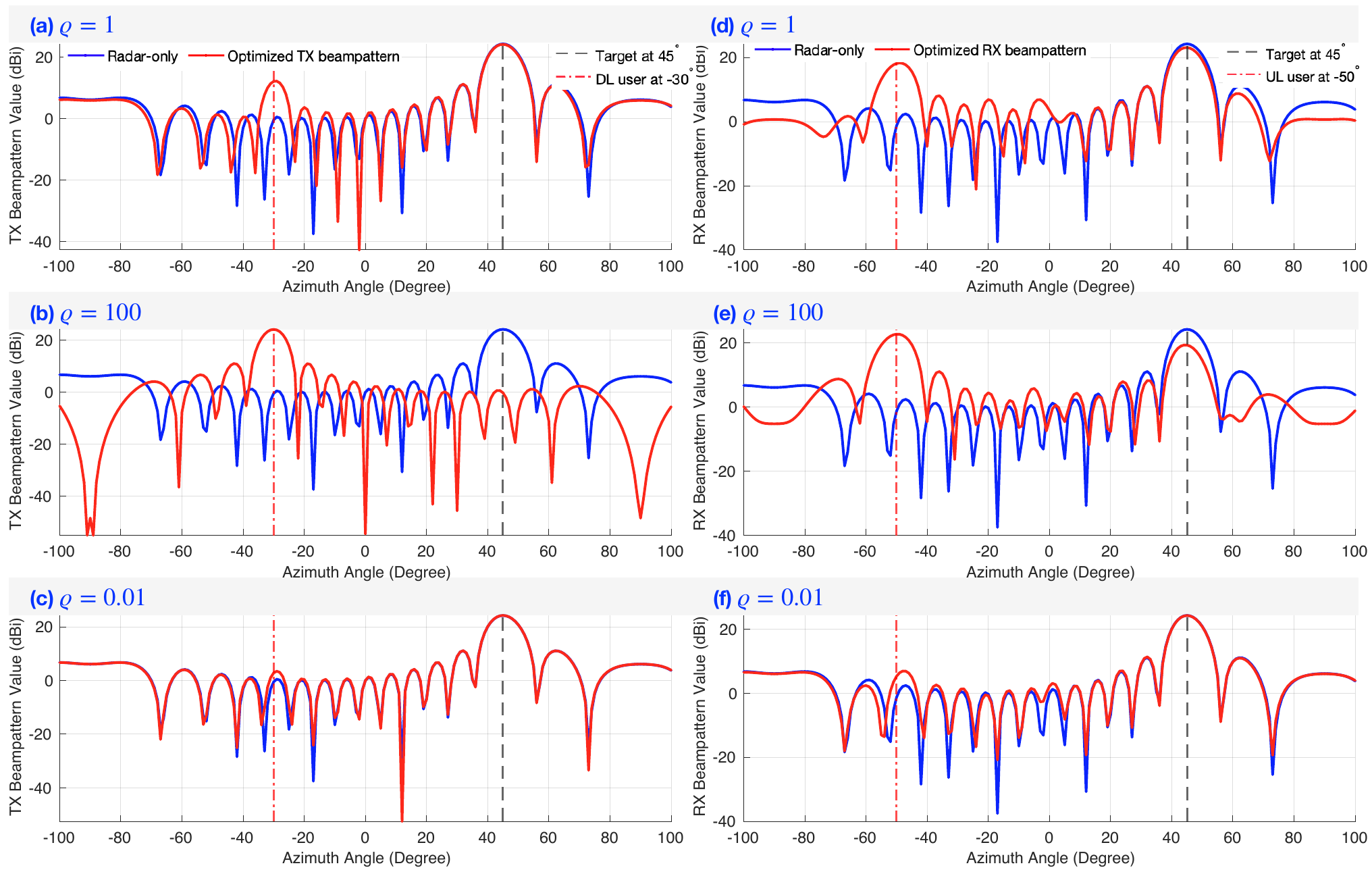}
		\caption{\textbf{Left panel}: Transmit beampattern, \textbf{Right panel}: Receive beampattern of the ISAC FD system with a effective cancellation of 60 dB SI, where $N_\mathrm{t}=16$, target at $\theta_\mathrm{r}=45^\circ$, and downlink user at $\theta_\mathrm{d}=-30^\circ$: (a, d) $\varrho =1$; (b, e) $\varrho =100$; (c, f) $\varrho =0.01$.}
		\label{fig:txrx}
	\end{center}
\end{figure*}

Fig \ref{fig:txrx} illustrate examples of the transmit and receive beampatterns with varying priority parameters $\varrho$ with effective cancellation of $60$dB residual SI. As shown in Figs. \ref{fig:txrx}a and \ref{fig:txrx}d, when $\varrho=1$, the transmit beamformer focuses the transmit power towards the radar target and the downlink user directions. Meanwhile, the receive beamformer concentrates on the uplink user and radar target. According to Figs. \ref{fig:txrx}b and \ref{fig:txrx}e, the transmit and receive beams focus more on the downlink user and uplink user compared with the transmit and receive beampatterns of $\varrho=1$, respectively, due to a higher priority on the communication that is $\varrho = 100$. When $\varrho = 0.01$, the radar function has higher priority, thus the transmit and receive beams concentrate more on the target in comparison with the transmit and receive beampatterns of $\varrho=1$. Additionally, the residual SI power level $P_\mathrm{res,[dB]}$ for the cases that $\varrho = 1, 100, 0.01$ are $-130.97$dB, $-197.37$dB, and $-108.48$dB, respectively. Hence, the SI is effectively suppressed. 


\subsection{Sum-rate Performance}
\label{ssec:sumrate}
\begin{figure}[t]
	\begin{center}
		\includegraphics[width =0.45\textwidth]{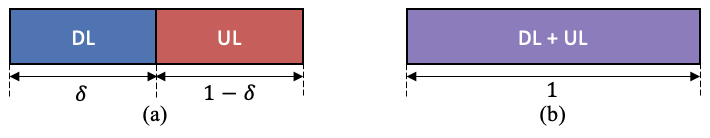}
		\caption{Frame structures of (a) half-duplex and (b) FD systems.}
		\label{fig:hd_frame}
	\end{center}
\end{figure}

\begin{figure}[t]
	\begin{center}
		\includegraphics[width =0.45\textwidth]{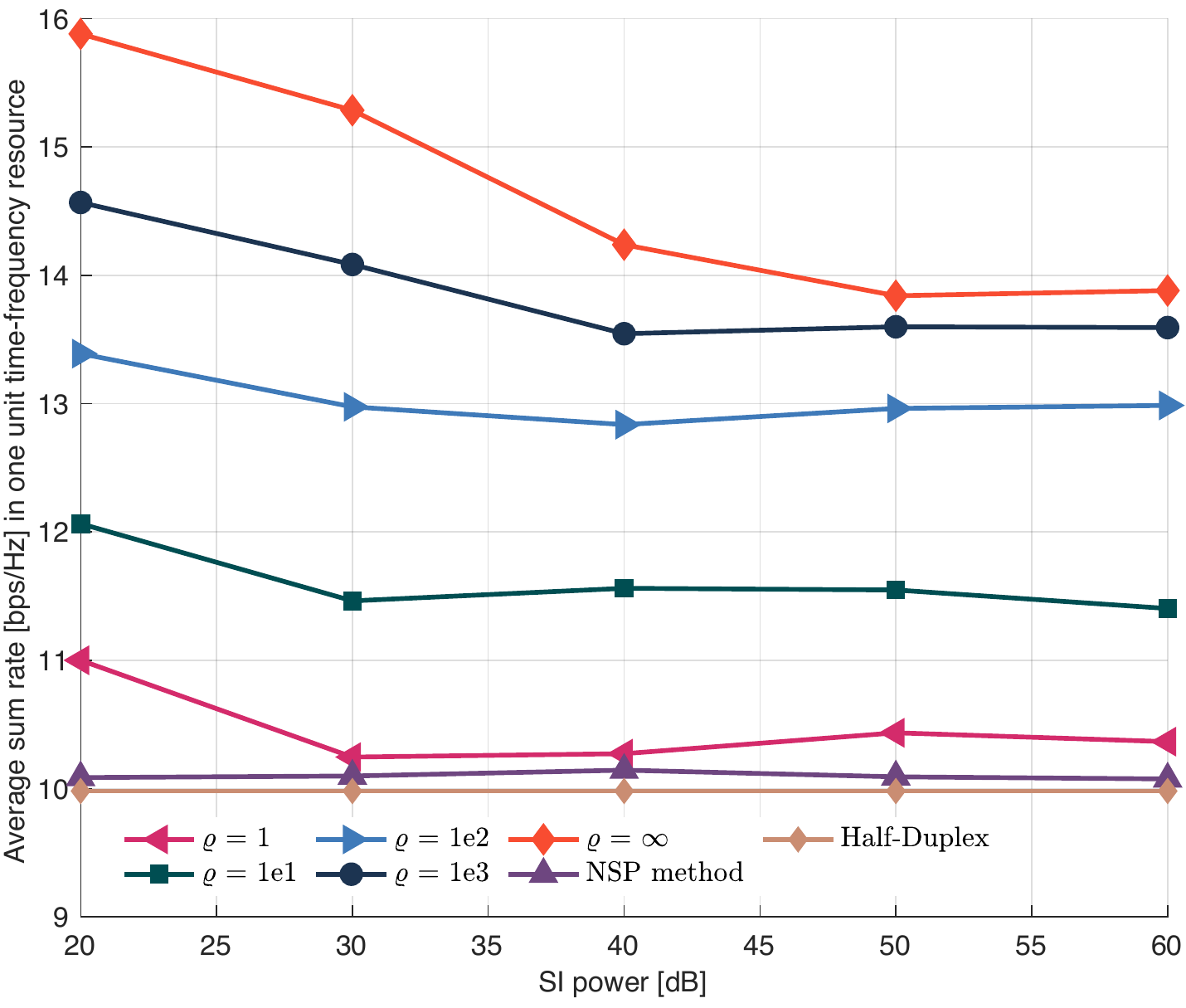}
		\caption{Average sum-rate in one unit time-frequency resource versus varying level of SI power.}
		\label{fig:srvssi}
	\end{center}
\end{figure}
In half-duplex systems, a fraction, $\delta$, of the (time-frequency) resources are allocated for uplink and the remaining for downlink, while for FD, all the resources are allocated simultaneously for both uplink and downlink, as shown in Fig. \ref{fig:hd_frame}.
The sum-rate in one unit time-frequency resource for the half-duplex system is defined as $R_\mathrm{hd,[unit]}=\delta R_\mathrm{dl} + (1-\delta) R_\mathrm{ul}$, and sum-rate for the FD system is defined as $R_\mathrm{fd,[unit]}= R_\mathrm{dl} + R_\mathrm{ul}$. In our simulation, $\delta$ is set as $0.5$, which means equal time-frequency resource allocation for both uplink and downlink.

The sum-rate performance with respect to SI power level with varying priority parameter $\varrho$ is shown in Fig. \ref{fig:srvssi}. When $\varrho$ increases, the sum-rate increases correspondingly. Specifically, the upper bound is given by the sum-rate of the FD communication functions (i.e., $\varrho = \infty$). The lower bound is given by the sum-rate in one unit time-frequency resource of the half-duplex system.  With increasing SI power, the sum-rate in unit time-frequency resource first decreases and then tends to be flat thanks to the effective suppression of the residual SI.

\subsection{Sensing Performance}
\label{ssec:radaresti}
For range-Doppler sensing, we explicitly consider the signal stream in (\ref{eq:signalbs}) in discrete time, ignoring the residual SI. The resulting signal can then be represented as
\begin{equation}
\begin{aligned}
\hat{s}_u[n, m] &= \mathbf{w}^H \mathbf{H}_\mathrm{u} \boldsymbol{\omega}_\mathrm{u} s_\mathrm{u}[n, m] \\
&+ \eta_\mathrm{r} e^{j 2 \pi (f_\mathrm{d} T_\mathrm{s}) m } \mathbf{w}^H \mathbf{a}(\theta_\mathrm{r}) \mathbf{b}^\top(\theta_\mathrm{r}) \mathbf{p} s_\mathrm{d}[n - i_\tau,m ] \\
&+ \mathbf{w}^H\mathbf{n}_\mathrm{u}[n, m], \\
& \quad (n = 0,\cdots,N; m = 0, \cdots, M-1), 
\end{aligned}
\label{eq:ybsdis}
\end{equation}
where the indices $m$ and $n$ respectively capture the slow and fast time-scales commonly assumed in range-Doppler processing\footnote{This framework results in a decoupling between the effects of the delay and Doppler shifts in (\ref{eq:ybsdis}), which is a reasonable assumption when the Doppler frequency ($320$Hz in our case) is much smaller than the signal bandwidth (20MHz in our case).}, and $i_\tau \in \mathbb{Z}$ is the round-trip delay of the radar echo (also known as the target range bin). The signal stream, $s_d[n, m]:=s_d[n + mM]$ (and likewise, with $s_u[n, m]$) can be viewed as a concatenation of $M$ blocks, with each block comprising $N$ symbols.

The range bin, $i_\tau$, is estimated from (\ref{eq:ybsdis}) by applying the matched filter w.r.t $s_d[n,m]$ along the fast-time axis. 
In \cite{aditya2022sensing}, it was shown that the cross-correlation function between $s_u[\cdot,m]$ and $s_d[\cdot,m]$, and the autocorrelation function of $s_{\rm d}[\cdot,m]$ asymptotically converged to the all-zero function and $\delta[\cdot]$, respectively, for large $N$.
Hence, the uplink data stream has negligible impact on the range-Doppler sensing performance. Subsequently, we perform an $M$-point DFT operation, resulting in the range-Doppler map.



\begin{figure}[tb]
	\begin{center}
		\includegraphics[width =0.5\textwidth]{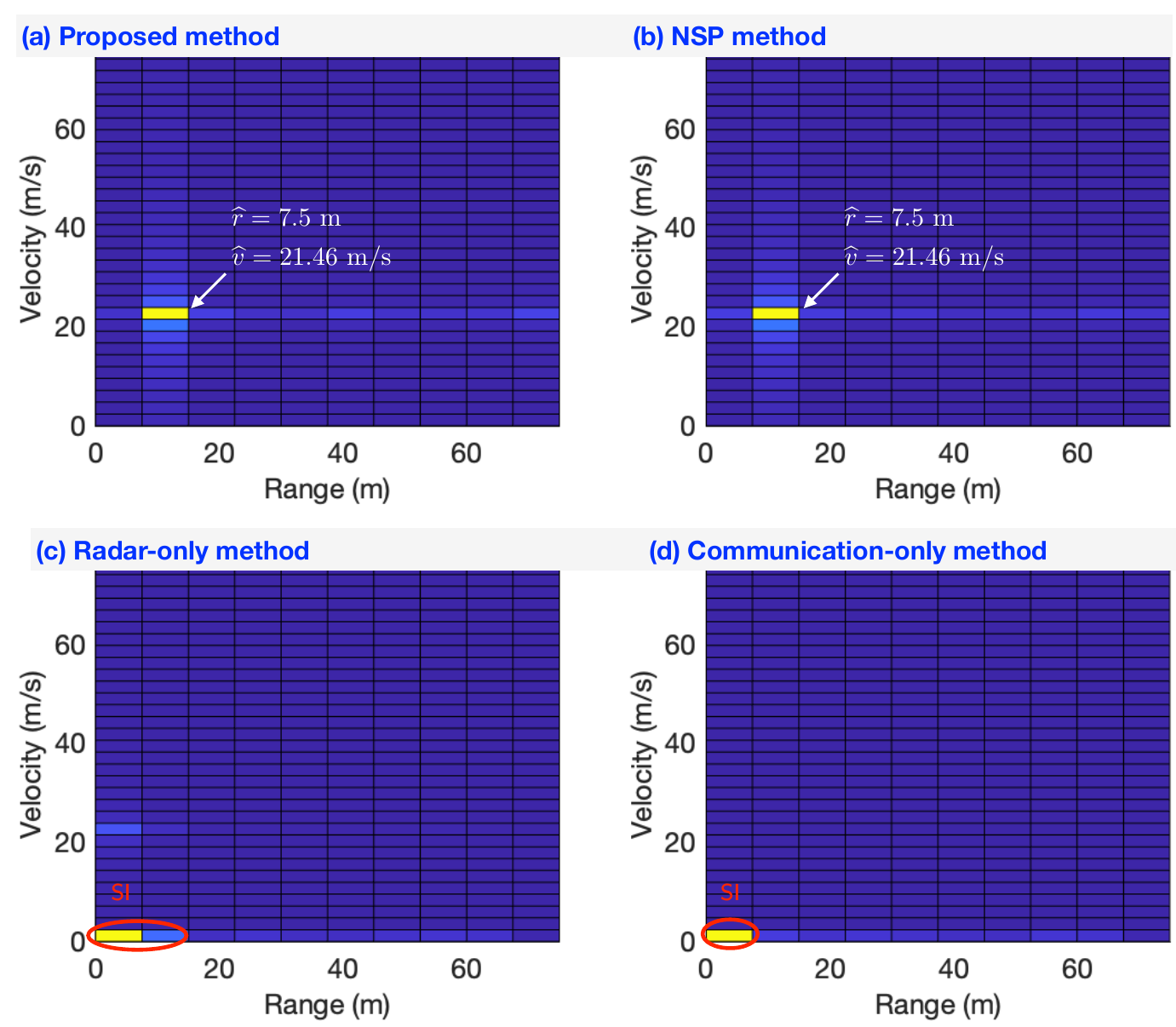}
		\caption{Examples of range and velocity estimation based on the FD ISAC system with $60$ dB SI with $N_\mathrm{t} = 16, N_\mathrm{r} = 16$, $M=512$ blocks, and $N=1024$ symbols.}
		\label{fig:dopplerrange}
	\end{center}
\end{figure}

With respect to angular domain sensing, since we implicitly assume that the target is in the vicinity of the direction in which $\mathbf{w}$ is "pointing" towards (i.e., the radar operates in \emph{track mode}), we are more interested in the interference suppression capability of the latter from an unwanted direction, $\theta$, captured by the following metric:
\begin{equation}
    \begin{aligned}
    P(\theta) & = | \mathbf{w}^H (\mathbf{a}(\theta) + \mathbf{H}_\mathrm{si} \mathbf{p}  )|^2.
    \end{aligned}
\end{equation}
Consequently, the AoA estimate is given by $\arg \max_\theta P(\theta)$.

For comparison, we consider the NSP method, radar-only method, and communication-only method. Specifically, the TX and RX beamformers for the radar-only method are given by $\mathbf{p}_\mathrm{r} = \mathbf{b}(\theta_\mathrm{r})$, and $\mathbf{w}_\mathrm{r} = \mathbf{a}(\theta_\mathrm{r})$, respectively. Similarly, the TX and RX beamformers for the communication-only method are given by $\mathbf{p}_\mathrm{u} = \mathbf{b}(\theta_\mathrm{u})$, and $\mathbf{w}_\mathrm{u} = \mathbf{a}(\theta_\mathrm{u})$, respectively. The priority parameter $\varrho$ is set to $1$. The range-velocity map is shown in Fig. \ref{fig:dopplerrange}. We observe that Algorithm \ref{alg:alg1} has effective residual SI suppression compared with methods that do not consider SI suppression (i.e., the radar-only and communications-only methods), where the desired radar echo drowns in the SI, which can be seen as a strong signal with $v=0\text{m/s}, r=0\text{m}$. Additionally, the radar-only and communication-only methods both have interference around the original point due to the unsuppressed residual SI.

\begin{figure}[tb]
	\begin{center}
		\includegraphics[width =0.5\textwidth]{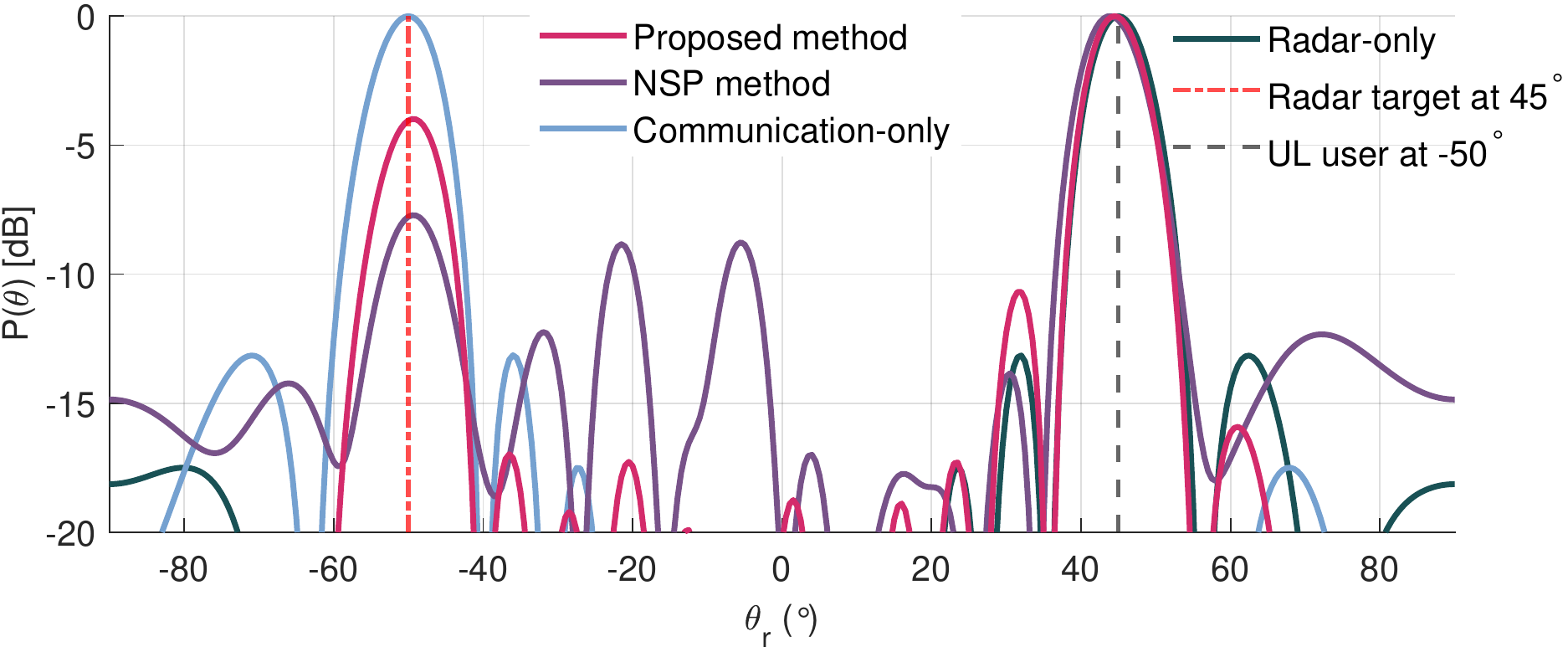}
		\caption{An example of angle estimation for a simulated scenario with a target at $\theta_\mathrm{r}=45^\circ$, a uplink communication user at $\theta_\mathrm{u}=-50^\circ$ with antennas $N_\mathrm{t}=16$ and $60$ dB SI power.}
		\label{fig:estangle}
	\end{center}
\end{figure}

Fig. \ref{fig:estangle} contains plots of the output power, $P(\theta)$, of different methods. We observe that the proposed method has a close angle detection output power, $P(\theta)$, at the target direction but a distinct output power at the uplink user direction, when compared with the radar-only method. The proposed method also has a similar output power compared with communication-only method. Additionally, the proposed method has fewer interference compared with the NSP method, especially around the angle $0^\circ$, where the SI locates. 

\section{Conclusion}
\label{sec:con}
In this paper, we design the transmit and receive beamformer $\mathbf{p}, \mathbf{w}$ at the transceiver, precoder $\boldsymbol{\omega}_\mathrm{u}$ at the uplink user, combiner $\mathbf{u}_\mathrm{d}$ at the downlink user to simultaneously maximize the uplink and downlink rate, the transmit and receive radar beampattern power at the target, and suppress the residual SI. In the objective function, TX and RX beampattern gains are used as the radar metric, and the uplink and downlink rates are used as the communication metric. With the aid of the equivalence of the rate maximization and the MSE minimization, and penalty-based transformation, we use the BCD method to solve the optimization problem. Subsequently, we give a convergence analysis. Numerical results show that up to 60 dB residual SI in digital domain can be efficiently suppressed. Additionally, the optimized TX and RX beampatterns can probe at the desired target, uplink and downlink user directions with a satisfactory average sum-rate, a more accurate radar parameter estimation with regards to range, velocity, and angle, and outperforms the NSP beamformer design method, which validates the effectiveness of the proposed algorithm. For future studies, a more effective algorithm on suppressing SI is worthy investigating. In addition, an extended transceiver design for a more general multi-data transmission can be studied.  Moreover, it would be interesting to extend the SI cancellation technique in digital domain to full domain in view of more accurate residual SI model and saturation caused by analog SI.  

\bibliographystyle{IEEEtran_url}
\bibliography{IEEEabrv,references}

\end{document}